\documentclass[%
preprint,
english,
showkeys,
nofootinbib, 
nobibnotes,
amsmath,amssymb,
]{revtex4-2}

\usepackage{caption}
\usepackage{graphicx}
\usepackage{dcolumn}
\usepackage{bm}
\usepackage{float}
\usepackage{hyperref}

\usepackage{amsmath}
\usepackage{tikz}
\usetikzlibrary{external,positioning}

\usepackage{xr}
\usepackage{hyperref}

\usepackage{mathdots}
\usepackage{yhmath}
\usepackage{cancel}
\usepackage{color}
\usepackage{array}
\usepackage{multirow}
\usepackage{amssymb}
\usepackage{tabularx}
\usepackage{extarrows}
\usepackage{booktabs}
\usetikzlibrary{patterns}
\usetikzlibrary{shadows.blur}
\usetikzlibrary{shapes,arrows.meta}
\usepackage{cleveref}
\usepackage{upgreek}
\usepackage[official]{eurosym}

\usepackage{color}

\newcommand{\mumgr}{$\upmu$Manager}
\newcommand{\zmax}{z_\mathrm{focus}}
\newcommand{\zmark}{z_\mathrm{mark}}
\newcommand{\zbeads}{z_\mathrm{beads}}
\newcommand{\mum}{\qty{}{\micro\meter}}
\newcommand{\ecoli}{\textit{E. coli}}

\newcommand{\NA}{N\!A}

\usepackage[utf8]{inputenc} 
\usepackage[english]{babel}
\usepackage{algorithm}
\usepackage{algpseudocode}
\usepackage{siunitx}
\usepackage{subcaption}
\usepackage[export]{adjustbox}

\makeatletter
\def\grd@save@target#1{%
  \def\grd@target{#1}}
\def\grd@save@start#1{%
  \def\grd@start{#1}}
\tikzset{
  grid with coordinates/.style={
    to path={%
      \pgfextra{%
        \edef\grd@@target{(\tikztotarget)}%
        \tikz@scan@one@point\grd@save@target\grd@@target\relax
        \edef\grd@@start{(\tikztostart)}%
        \tikz@scan@one@point\grd@save@start\grd@@start\relax
        \draw[minor help lines] (\tikztostart) grid (\tikztotarget);
        \draw[major help lines] (\tikztostart) grid (\tikztotarget);
        \grd@start
        \pgfmathsetmacro{\grd@xa}{\the\pgf@x/1cm}
        \pgfmathsetmacro{\grd@ya}{\the\pgf@y/1cm}
        \grd@target
        \pgfmathsetmacro{\grd@xb}{\the\pgf@x/1cm}
        \pgfmathsetmacro{\grd@yb}{\the\pgf@y/1cm}
        \pgfmathsetmacro{\grd@xc}{\grd@xa + \pgfkeysvalueof{/tikz/grid with coordinates/major step}}
        \pgfmathsetmacro{\grd@yc}{\grd@ya + \pgfkeysvalueof{/tikz/grid with coordinates/major step}}
        \foreach \x in {\grd@xa,\grd@xc,...,\grd@xb}
        \node[anchor=north] at (\x,\grd@ya) {\pgfmathprintnumber{\x}};
        \foreach \y in {\grd@ya,\grd@yc,...,\grd@yb}
        \node[anchor=east] at (\grd@xa,\y) {\pgfmathprintnumber{\y}};
      }
    }
  },
  minor help lines/.style={
    help lines,
    step=\pgfkeysvalueof{/tikz/grid with coordinates/minor step}
  },
  major help lines/.style={
    help lines,
    line width=\pgfkeysvalueof{/tikz/grid with coordinates/major line width},
    step=\pgfkeysvalueof{/tikz/grid with coordinates/major step}
  },
  grid with coordinates/.cd,
  minor step/.initial=.2,
  major step/.initial=1,
  major line width/.initial=2pt,
}
\makeatother

\begin{document}

\title{An Affordable Fiducial Marker Strategy for Reliable Autofocus in Long-Term Live Microscopy}
\author{Ilyas Djafer-Cherif}
\email{ilyas.djafer-cherif@ichf.edu.pl}
\affiliation{Institute of Physical Chemistry, Polish Academy of Sciences}
\author{Bartlomiej Waclaw}%
\email{bwaclaw@gmail.com}
\affiliation{Institute of Physical Chemistry, Polish Academy of Sciences.\\ Present affiliation: Ogibiotec, Pentland Science Park, Bush Loan, EH26 0PZ, United Kingdom}

\date{\today}

\begin{abstract}
Long-term time-lapse imaging of biological samples requires correcting for focal drift, which would otherwise gradually push the sample out of focus. We present a software-based method that eliminates this time-dependent blur using only a motorized Z-drive, with no additional hardware. The method relies on imaging marks made on the side of the coverslip opposite to the sample. We provide a Beanshell script implementation, evaluate its performance across multiple objectives, and benchmark it against a hardware autofocus system, finding comparable results. Finally, we demonstrate its effectiveness in live imaging of growing bacterial colonies.
\end{abstract}

\keywords{Software-based focus system, low-cost autofocus}

\maketitle{}
{}

\section{\label{sec:intro} Introduction}
Time-lapse wide-field microscopy is a widely used method to image live biological samples that are thin compared to the depth of field of the microscope objective. Examples of experimental setups when optical sectioning is not required range from small colonies of microorganisms on agarose pads \cite{kennard_individuality_2016,duvernoy_asymmetric_2018}, to bacteria in a mother machine \cite{taheri-araghi_cell-size_2015,bakshi_tracking_2021} or a micro-chemostat \cite{balagadde_long-term_2005}, to animal cells in microfluidic channels \cite{wu_cell_2013}. A common issue affecting image quality in such experiments is the variability of the relative position of the microscope objective lens and the sample due to thermal drift, sample aging, and mechanical distortion caused by pressure changes during the operation of microfluidics chips \cite{drift-sources}. The resulting movement of the imaging focal plane must be corrected, otherwise the image will go out of focus. Since imaging can last many days, manual correction is not practical, and some sort of automated focus adjustment is usually required.

Most research-grade microscopes can be equipped with a motorized stage, a motorized objective turret, or a piezo objective scanner which can move the objective in the Z axis (towards/away from the sample). Automated focus correction can be implemented either in the imaging software or as a separate hardware module. Although software-based methods have the advantage of not requiring additional hardware, their performance suffers when the sample does not have sharp, well-defined features or when the features change over time. Hardware-based systems generally use techniques based on monitoring the distance between the objective and a surface of the sample with a sharp change of the refractive index; this usually coincides with the surface of a glass slide or the bottom of a microplate or a culture dish. These systems work well in many scenarios but are expensive and require compatible objectives. Moreover, hardware methods are not entirely immune to problems affecting software-based methods, such as significant changes of the optical properties of the sample, for example, due to biological growth, air bubbles in microfluidic channels, and sample ageing.

Here we propose an "imPerfect Focus System" (iPFS) - a method which uses a zero-cost fiducial marker \cite{colomb_estimation_2017}  associated with a simple, yet effective,  software-based method. The only non-software requirement is a computer-controlled hardware for moving the objective in the optical (Z) axis. Despite the name, we will demonstrate that the method can be as accurate as the hardware-based Nikon Perfect Focus System. 
Our method uses a computer algorithm to track user-made markers on one of the rigid surfaces of the sample (glass slide, cover slip, or plastic dish). Changes in the Z coordinate of this reference surface are then translated into offsets applied to imaging positions.
Markers can be added using a Sharpie, or by utilizing natural imperfections such as dust, grease, or crystal deposits already present on the surface of microscopic slides. This method is compatible with both transmitted-light and fluorescence imaging, allows independent focusing across multiple XY positions (fields of view), and does not interfere with optical imaging as long as at least 10–20 seconds are permitted between successive measurements.

\subsection{Overview of software-based focus methods}
There are many software-based autofocus methods that rely on numerical techniques to objectively quantify image sharpness, which can then be used to bring the sample into focus. Earlier approaches estimate sharpness by applying a mathematical function that assigns a numerical 'sharpness' value to the image, often using filtered outputs such as when using the Laplacian or Sobel operators. Other approaches looking at the power spectrum of the image \cite{firestoneComparisonAutofocusMethods1991} or leverage how information is encoded in the spectral domain the Discrete Cosine Transform which is used in JPEG compression \cite{kristanBayesspectralentropybasedMeasureCamera2006}, some even define sharpness inferred from how human can perceive the minimal amount of "Just noticeable blur" \cite{ferzliNoReferenceObjectiveImage2009}.
However, most of these methods assume that the sample contains sufficient structural features for the sharpness metric to exhibit a clear, rapidly decaying peak away from the focal plane—in other words, a well-defined local maximum. Additionally, the sharpness peak must be well separated from other local maxima, otherwise the algorithm may randomly jump between these maxima. These assumptions may not hold for certain sample types, such as thick biological specimens. Moreover, for such samples one may be interested in imaging a different plane to that for which the sharpness function is maximized.

The sharpness function may also be rugged, i.e., exhibit many local peaks caused by the sample structure and intrinsic optical and camera noise, and while some methods claim to alleviate this problem \cite{jiaEnhancedMountainClimbing2025}, they are generally only applicable to samples exhibiting a well-pronounced single global sharpness maximum. 

More recent machine learning approaches replace traditional filters with deep convolutional networks, which can provide improved generalization \cite{shajkofciDeepFocusFewShotMicroscope2020}, or even reconstruct a synthetic in-focus image from a single real defocused image \cite{luoSingleShotAutofocusingMicroscopy2021}. However, such approaches generally rely on the imaged specimen being similar to what the algorithms have been trained on, and may exhibit reduced performance if the real sample type is not adequately represented in the training dataset.

Growth presents another challenge: as a sample evolves over time, its sharpness characteristics can change significantly. For example, \textit{Escherichia coli} colonies may transition from one-dimensional to two-dimensional structures \cite{grantRoleMechanicalForces2014}, effectively increasing sample thickness. This significantly changes the sharpness function, often breaking the two assumptions mentioned earlier (a sharp global maximum isolated from other local maxima).

\subsection{\label{sec:overview}Overview of hardware-based focus methods}

We discuss here two hardware-based focus systems: Nikon's Perfect Focus System \cite{ref_PFS} and an open-hardware alternative pgFocus \cite{pgfocus_website} . Systems used by other microscope manufacturers (Zeiss, Leica, Revvity) and do-it-yourself alternatives generally follow similar principles to these two systems. 

\subsubsection{Commercial systems: Nikon \textit{Perfect} Focus System 4}\label{sec:pfs}

This hardware-based system will serve as our reference for benchmarking. The system relies on partial reflection of infrared light at the sample’s air–glass or water–glass interface. Light from an infrared LED passes through a narrow slit and is directed through the imaging objective onto the sample, where it is partially reflected back to the objective and captured by a linear CCD sensor. The position of the resulting intensity peak on the sensor indicates the distance between the objective and the reflective surface. An electronic controller maintains this distance fixed by adjusting the position of the objective along the Z axis. An additional lens inserted in the optical path of the infrared beam allows the user to introduce an offset between the reflective surface and the imaging plane, enabling imaging objects above or below the reflective surface. A detailed explanation is provided on Nikon's website \cite{ref_PFS2}. 

The system works in real time (200 measurements/s) and its claimed accuracy is $1/3$ of the objective focal depth \cite{ref_PFS2}. 
However, actual performance depends on both the sample being imaged and the type of objective used. With water- or oil-immersion objectives, the primary reflection occurs at the glass-water interface, near the location of most biological samples. In contrast, for dry objectives the primary reflection occurs at the air-glass or air-transparent bottom interface, which can be hundreds of micrometres below the sample, depending of the substrate thickness. If the substrate thickness is uneven - a common issue for plastic-bottom plates - manual correction may be needed independently for each field of view. The system is also optimised for substrates with a refractive index close to that of standard microscopic cover slips ($n\approx 1.5$) and may perform poorly with some plastic-bottom plates. Moreover, certain objectives are incompatible with the PFS system. Another limitation is the lack of an open-source API for the PFS, which makes writing custom acquisition software difficult.

\subsubsection{pgFocus}\label{sec:pgfocus}
The PFS and similar systems can represent a significant portion of the cost of an already expensive automated microscope. Moreover, these solutions may not perform optimally for certain applications. Consequently, open-source alternatives have been developed. One example is pgFocus ("pg" stands for \emph{pretty good}), which can be assembled from readily available components. The schematics and component part numbers are provided in a Git repository \cite{pgfocus_hw}, and the control software is available as a \mumgr{} plugin. 

In contrast to the PFS described above, this system measures reflected light arising from total internal reflection (TIR) at the interface. It requires a high numerical aperture (NA) objective and is therefore compatible only with water- or oil-immersion objectives. The system has been reported to achieve higher accuracy than the PFS \cite{pgfocus_table} and can operate continuously during 3D (z-stack) acquisition, a capability essential for what it has been designed: total interference reflection fluorescence (TIRF, not supported by the PFS). 

\section{Imperfect Focus System}
The \emph{Imperfect Focus System} (iPFS) proposed here integrates two concepts previously implemented separately in the hardware- and software-based systems described above: (i) finding a rigid surface that moves with the sample, (ii) using the same camera and objective employed for imaging the sample. iPFS requires no additional hardware beyond a PC-controllable Z stage and is compatible with any objective.

\begin{figure*}
    \centering\includegraphics[width=\linewidth]{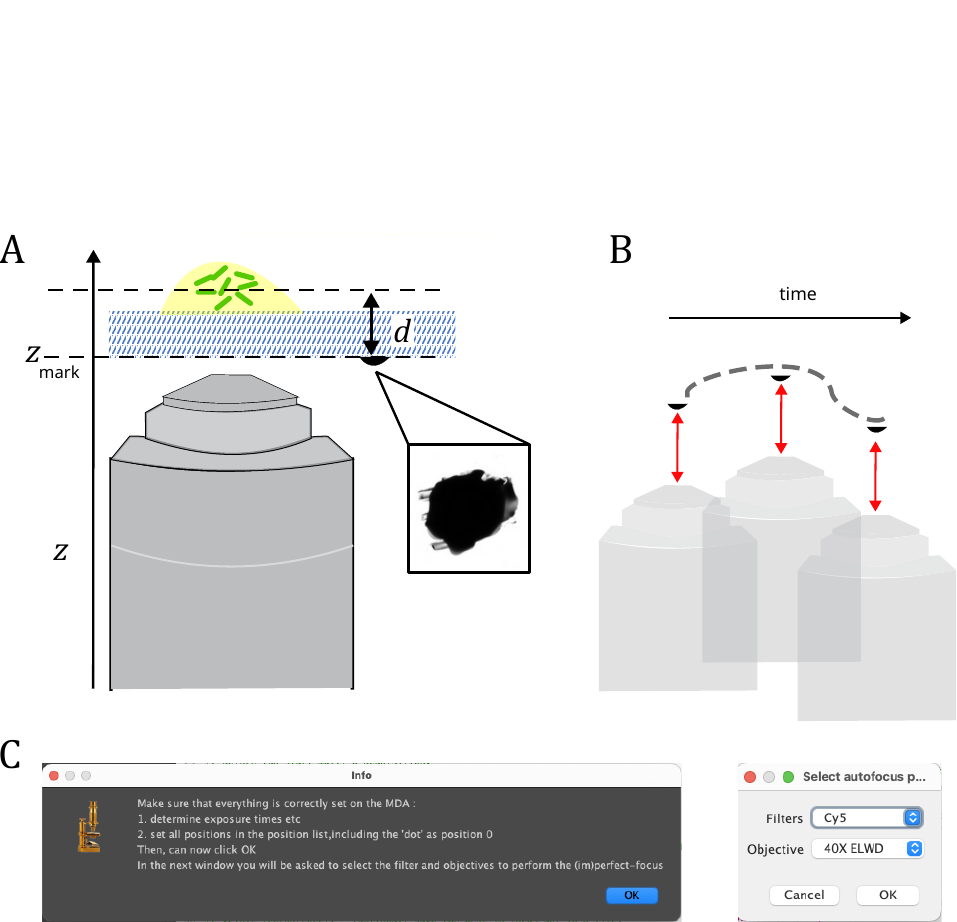}\\
    \caption{Schematic of iPFS operation. (A) A marker on the external surface of the microscope slide is kept in focus, serving as a reference for all distances in the Z axis. (B) The algorithm maintains sample focus by tracking the marker over time. (C) The \mumgr{} \emph{iPFS} script interface. 
    }
    \label{fig:schem}
\end{figure*}

The operating principle is inspired by hardware-based PFS: we detect a "rigid surface" of the sample - typically the bottom surface of the coverslip or plastic sample container - and adjust all imaging positions to follow its movement, compensating axial drift (Figure \ref{fig:schem}A-B). Unlike PFS, we track optically thin, high-contrast markers on the transparent surface of the sample to monitor Z-axis motion during imaging. Because the marker and the sample are part of the same rigid body, the detected movement of the marker can be used to offset the imaging plane, compensating for sample drift over time.

The markers are imaged in transmitted light (bright field or phase contrast) during pauses in sample imaging. The markers must be firmly attached to the transparent surface and exhibit sharp features such as edges or points that blur when moved out of the focal plane. We found that markers created by touching the glass slide very lightly with a marker pen work very well. However, naturally occurring marks such microscopic mineral deposits, micro-scratches and similar quatwo dimensional objects can also serve as markers, provided they remain stable over time. 

In what follows we shall discuss the algorithm and its proposed implementation as part of \mumgr{}'s Acquisition engine. 

\subsection{Algorithm\label{sec:algorithm}}
The algorithm runs continuously during pauses in image acquisition, enhancing stability and reducing the risk of losing focus in the event of sudden sample movement.

A single iteration of the algorithm consists of three steps: (i) move the stage to the previously recorded position of the marker (i.e., move both the XY and Z stages) used by the autofocus routine, (ii) acquire a 3D image stack around the Z coordinate corresponding to the sharpest image of the marker from the previous step, and (iii) update the marker's Z coordinate to the position of the sharpest image in the stack. The final step involves calculating a "sharpness function" for each image. Some possible choices of the sharpness function will be discussed in the next section. The following pseudo-code illustrates these steps.

\begin{algorithm}[H] 
\caption{The \emph{Imperfect Focus} algorithm}\label{alg:ipfs}
\begin{algorithmic} 
\State $(x_0,y_0,z_0)$ \Comment{Initial position of the mark}
\State $(x_{i},y_{i},z_{i})_{i \in [1,n]}$ \Comment{Sample positions to be imaged}
\State $z_{\rm range}$ \Comment{scan range}
\State $\delta z$ \Comment{step size}
\State $\zmax  = z_0$ \Comment{We assume the mark is initially in focus}
\While{True}
\If{EnoughTimeUntilNextAcquisition()}
\For{ $ \zmax - z_{\rm range} \leq z \leq \zmax + z_{\rm range} $ }
    \State acquire image at $(x_0,t_0,z)$
   \State compute $\mathrm{Sharpness}(z)$
   \State $z = z + \delta z$
\EndFor
\State $\zmax = \max\left(\mathrm{Sharpness}(z) \right)_{ [ \zmax-z_{\rm range},\zmax+z_{\rm range} ]} $
\Else
\State $\mathcal{O} = \zmax - z_{0} $ \Comment{Compute drift}
\State $z_0 = \zmax$ \Comment{Update mark position}
\State $\{z_i\} =  \{ z_i +  \mathcal{O}\} $ \Comment{Offset all imaging position}
\State ImageAllPositions($\{x_i,y_i,z_i\}$) \Comment{Actual image acquisition happens here}
\EndIf
\EndWhile
\end{algorithmic}
\end{algorithm}

We note we use a rather naive approach to finding the maximum of the sharpness function along the Z coordinate by scanning over a range of Z values, rather than using potentially more efficient 1D minimization routines. Despite its simplicity, this method ensures stability and convergence to the global maximum, while the number of function evaluations remains comparable to more advanced methods, given the relatively modest accuracy required ($0.1\mu$m or less). In addition, this approach allows one to move the objective always in the same direction, reducing hysteresis in the Z-stage drive mechanism and improving repeatability.

\subsection{Choice of the sharpness function}
The sharpness function takes a 2D image and returns a sharpness metric - a real number that reaches its maximum when the marker is exactly in the focal plane of the objective. Many edge-detecting filters followed by integration exhibit this property \cite{mirExtensiveEmpiricalEvaluation2014}.
Other options include normalised variance, entropy of the intensity histogram \cite{redondoAutofocusEvaluationBrightfield2012} and machine-learning based methods \cite{yangAssessingMicroscopeImage2018}.

Our criteria for selecting an appropriate sharpness metric were: (i) high sensitivity to defocusing when imaging marker-pen marks, (ii) robustness against camera noise and variations in image brightness, and (iii) ease of implementation in Beanshell, the Java-based scripting language used by \mumgr{}. We tested several different sharpness metrics before finally selecting the 3x3 so-called "Redondo" filter:
\begin{equation}
R = 
    \begin{bmatrix}
         0 & 1 & 0  \\
         -3   & 0 & 1  \\
         0 & 1 & 0 \\
    \end{bmatrix}.
    \label{eq:redondo_matrix}
\end{equation}
This particular choice of filter is not critical to the performance of the algorithm; see SI Figure \ref{fig:si_comparison_filters} for a comparison of different filters.
Image sharpness is obtained by convolving the matrix (\ref{eq:redondo_matrix}) with the image intensity matrix $I_{x,y}$, 
squaring all elements, and summing them up: 
\begin{equation}
  S = \sum_{x,y} \left( \sum_{i,j} R_{ij} I_{x+i,y+j} \right)^2.
\end{equation}
In what follows, we will refer to the numerical value of $S$ defined by the above equation as the sharpness metric, or the sharpness function.

\subsection{Implementation in \mumgr{}}
To demonstrate the algorithm and evaluate its performance on real microscopy images, we implemented it as a Beanshell script for \mumgr{}, a popular open-source software for microscope controll \cite{micromanager}.
The script can be run directly from the Script Panel without any additional software and is fully integrated with \mumgr{}'s Multi-Dimensional Acquisition (MDA). After setting up the usual MDA settings (timepoints, Z-stacks, etc.) through the usual GUI, the user must (1) ensure that the first position in the \mumgr{} positon list contains the marker, (2) launch the script via Beanshell, and (3) select the objective and the channel used to image the marker (see Figure \ref{fig:schem}C).
The script then initializes all relevant variables, records the initial marker position, and goes idle. It wakes after each time step of live image acquisition, running continuously before returning to idle just prior to the next acquisition.

\begin{figure*}
    \centering\includegraphics[width=\linewidth]{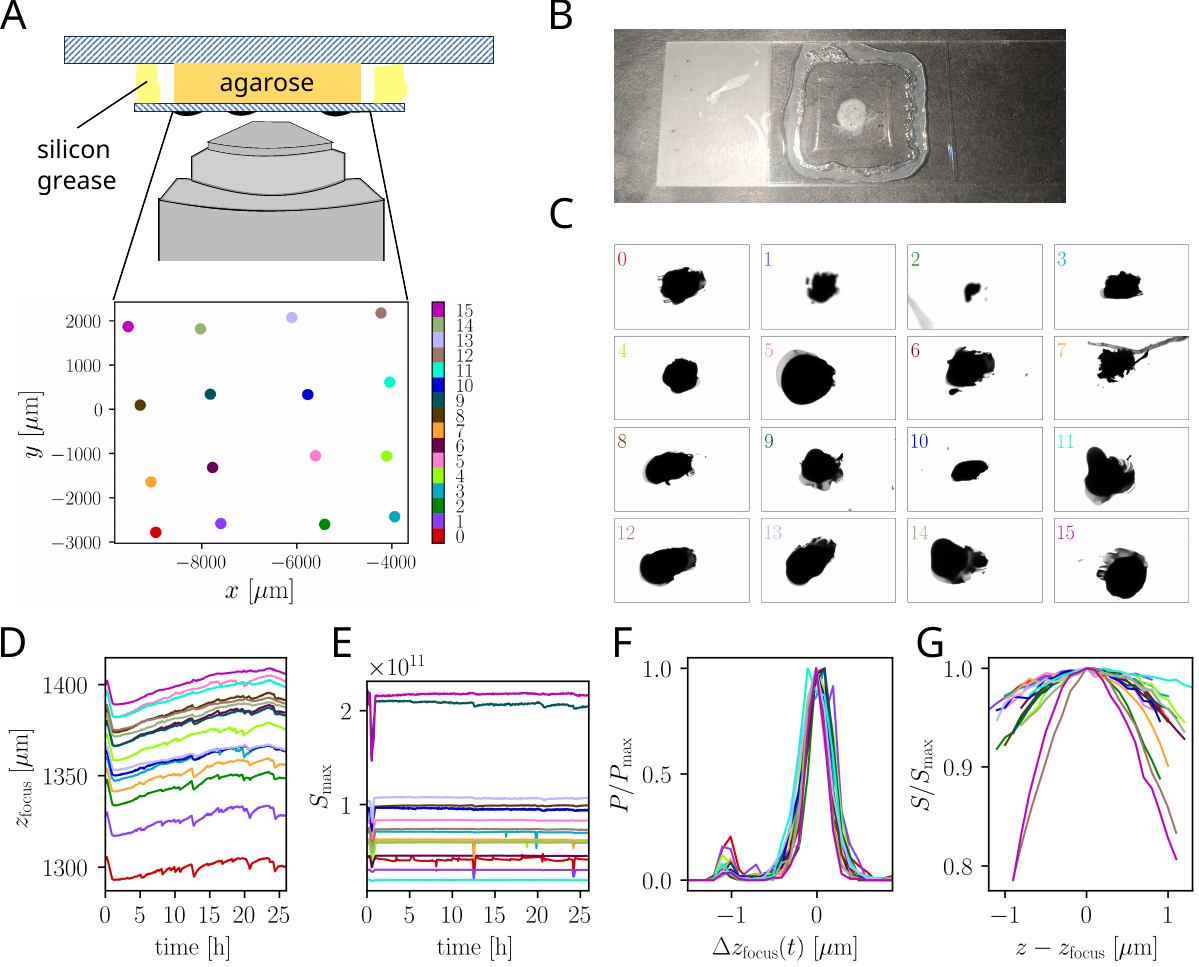}
    \caption{Experimental setup used to characterise the stability of the iPFS algorithm. (A) Schematic of the setup. The 16 marker positions used to evaluate stability are shown as coloured dots. (B) Photograph of the actual sample. (C) Images of individual markers.
   (D) Sharpness-maximizing Z positions versus time, for all 16 markers (20x magnification). (E) Corresponding sharpness metric as a function of time.  (F) Normalized probability distribution of changes in marker position, showing the deviation between consecutive measurements: $\zmax(t+3\ \mathrm{min}) - \zmax(t) = \Delta \zmax$. (G) Normalized sharpness curves for all markers at the end of the experiment. }
    \label{fig:16_dots}
\end{figure*}

\section{Performance assessment of the \emph{Imperfect Focus System}}\label{sec:performance}
We assessed the performance of our approach in a range of tests involving a sample slide containing microscopic ($\leq \qty{5}{\micro\meter}$) fluorescent beads sandwiched between a glass cover slip and an agarose pad
as well as sixteen marks made on the external surface of the coverslip in an area of approx. $15$ mm $\times15$ mm
(Figures \ref{fig:16_dots}A-B and  \ref{fig:beads}A). The beads served as a model specimen for long-term imaging. Their size was comparable to that of bacterial cells, making clusters of beads representative of small bacterial microcolonies. This setup poses a challenge for live imaging, as high-magnification, high-NA objectives are required, and even minor axial drift can severely degrade image quality. By using static beads instead of live bacteria, we were able to evaluate the long-term stability of the algorithm under conditions where image quality was affected almost exclusively by axial drift, with no significant sample-induced changes over time.

\subsection{Testing the ability of iPFS to track the movement of the sample}\label{ssec:prelim_tests}
We first evaluated how effectively the algorithm could track the axial movement of the sample and how its performance varied with differences in marker shape (Figure \ref{fig:16_dots}C). 

After mounting the sample on the XY stage and setting the microscope incubator to $\qty{37}{\celsius}$, all 16 markers were imaged continuously for 25 hours at 3-minute intervals using a 20x objective. At each time point, the iPFS algorithm computed the sharpness function $S(z)$ of each marker by scanning a narrow range of $z$ values around the previously determined peak position $\zmax$ of the sharpness function, and then updated $\zmax$ to the new maximum of $S(z)$. Only the sharpest image for each marker was saved; no other images from the scan were recorded in this test. Because all markers were located on the same coverslip, they were expected to move together, such that their $z$ coordinates would change by the same amount in response to any sample movement.

Figure \ref{fig:16_dots}D-E shows the time-series of $\zmax(t)$ and the corresponding sharpness metric for each marker. As expected, relative changes in all $\zmax$ coordinates were highly correlated. 
The sharpness metric, while varying between the markers, remained relatively stable during the experiment, aside from occasional, short-lived fluctuations. Visual inspection of the recorded sharpest images of the markers showed that these fluctuations and the associated jumps in $\zmax$ can be attributed to (i) sudden sample movements caused by agarose shrinking and detaching from the coverslip, (ii) out-of-focus background changes, such as the growth of air bubbles in the agarose. 

Figure \ref{fig:16_dots}F shows histograms of changes in $\zmax$ between consecutive time points for all markers. Aside from occasional sudden jumps of approximately $1\mum$, the distributions are concentrated near zero and have similar widths across all 16 markers. This demonstrates that the algorithm can track all markers equally well, even though their sharpness functions have very different widths (Figure \ref{fig:16_dots}G).  

The small differences in marker movement visible in Figure \ref{fig:16_dots}E are likely due to slight bending of the coverslip caused by mechanical forces from ageing or drying agarose.

\subsection{Accuracy of iPFS}
Figure \ref{fig:16_dots} suggests that the algorithm can reliably track the surface with $\sim\mu$m accuracy when using a 20x objective, irrespective of the markers' shape. We expect tracking accuracy to depend on the depth-of-field (DOF) of the objective: objectives with a shallower DOF should increase the algorithm’s sensitivity, as small changes in the Z coordinate have a greater impact on image sharpness.
The DOF is in general the sum of wave and geometrical contributions \cite{Oldenbourg_Shribak}:
\begin{equation}
    d = \frac{\lambda n}{\NA^2} + \frac{n\,p}{M\cdot \NA} ,
\end{equation}
where $n$ is the refractive index of the imaging medium between the lens and the coverslip, $\NA$ is the numerical aperture of the objective, $\lambda$ is the wavelength, and $p$ the size of the smallest detail resolvable by the camera, equivalent to the physical pixel size of the camera's sensor. 
For our air objectives ($M,\NA=(20,0.5)$ or $(40,0.95)$, $n=1$) (cf. Methods, Sec. \ref{ssec:microscopy}), and $\lambda = 550$ nm, the DOF $d\approx\lambda n/(\NA)^2$ turns out to be \qty{2.2}{\um} (20x objective) and \qty{0.7}{\um} (40x objective). We compare this with the full width at half maximum (FWHM) of the distribution of algorithm's steps in Figure \ref{fig:16_dots}F. The obtained value $\approx \qty{0.25}{\um}$ is much less than the DOF, meaning that the sample is expected to remain in focus.

\begin{figure*}
    \centering
    \includegraphics[width=0.8\linewidth]{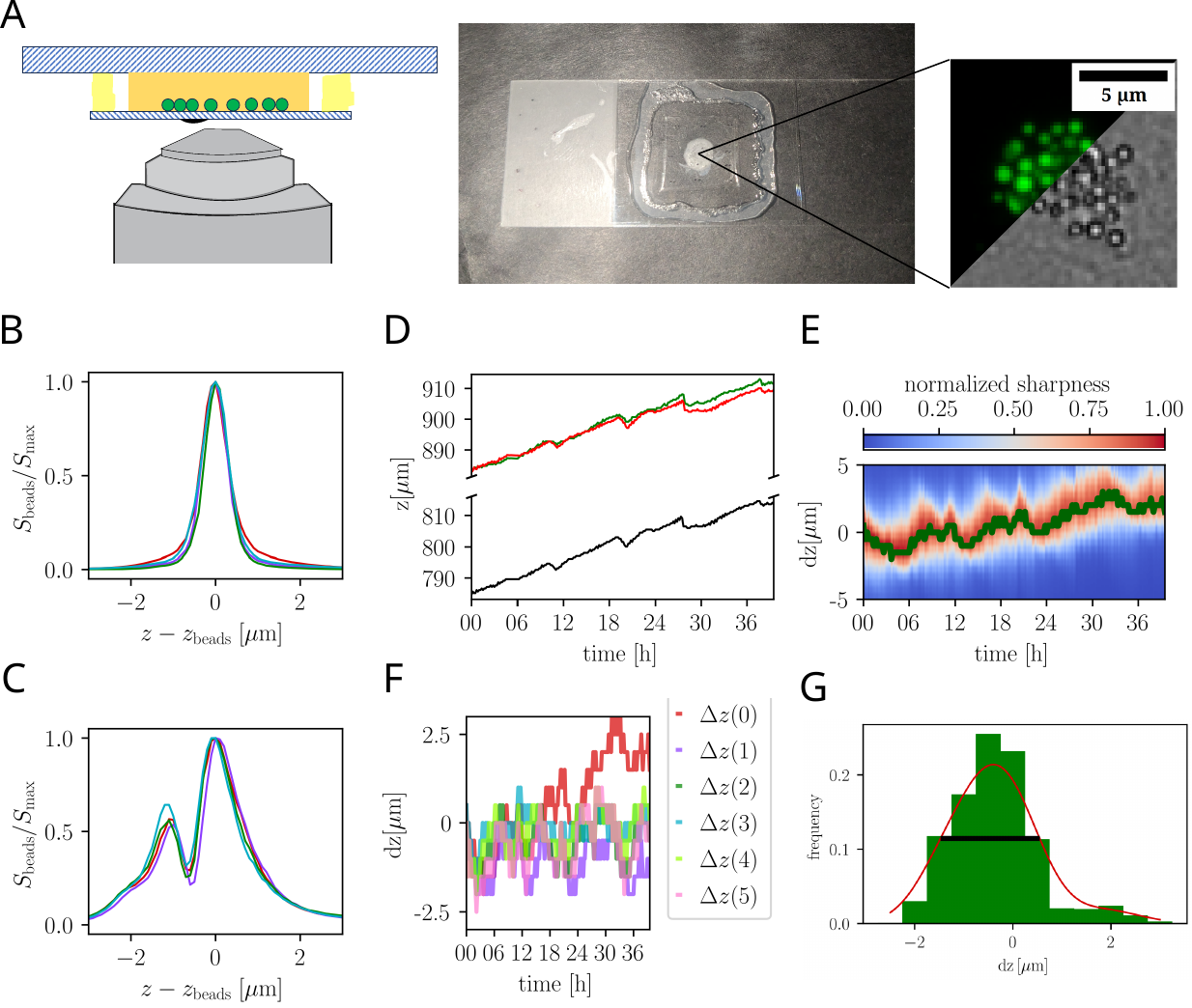}
    \caption{Imaging fluorescent beads using iPFS. 
    (A) Experimental setup. Viewed from the objective side, we see the following: coverslip (thin dashed rectangle) with a marker (black dot), fluorescence beads (green circles, not to scale), a 1 mm-thick agarose slab (orange rectangle), and the glass slide (thick dashed rectangle). The sample is sealed with silicone grease (yellow rectangles). An example image of the beads in both bright field and fluorescence is also shown.
    (B-C) Normalised sharpness of the beads along the z-direction (40x objective), at four different locations on the slide (curves in different colours), in fluorescence (B) and bright-field (C).
    See Figure S2 for the same plot for the 20x objective.
    (D) iPFS maintains focus despite sample movement (20x objective). Red line: $\zbeads$, the $z$ coordinate of maximum sharpness $S_{\rm beads}(z)$ for the beads at one position. Black line: $\zmark$, the position of the tracked marker. Green line: $\zmark$ offset by $d(0)$, the initial difference between the marker and the beads' sharpest image. Perfect drift compensation would make the green line overlap with the red line.
    (E) Sharpness function $S_{\rm beads}(\delta z + \zbeads(0) + (\zmark(t)-\zmark(0)))$, where $\delta z$ 
    is the offset from the corrected $z$ coordinate of the beads' sharpest image. Dark green line: offset $\Delta z$ of the sharpest image (maximum of $S(\delta z)$).
    (F) $\Delta z$ versus time for all six imaged locations. 
    (G) Histogram of $\Delta z$ for all six locations and all time points. 
    The FWHM (black line) has been obtained from a Kernel density estimate (red line) fitted to the histogram. }
    \label{fig:beads}
\end{figure*}

\subsection{Fluorescent beads as a test sample for benchmarking iPFS\label{sec:fluor_iPFS}}
To evaluate the algorithm’s ability to maintain focus on an actual sample, we imaged fluorescent beads from the same preparation used in the previous section, tracking one of the 16 markers with iPFS (Figure \ref{fig:beads}A).
Clusters of fluorescent beads resulted in a sharpness function with a single, narrow peak (FWHM $\approx \qty{0.9}{\um}$) when imaged in fluorescence mode (Figure \ref{fig:beads}B), and a broader, double-peaked profile in bright-field (Figure \ref{fig:beads}C, FWHM $\approx \qty{1.4}{\um}$); see SI Figure \ref{fig:fwhm} for the average profile and FWHM estimation. The fluorescence-based sharpness metric has a similar shape for different clusters of beads. Using a standard autofocus algorithm in bright-field mode on this sample could result in selecting either peak depending on the initial Z position, or cause the focus to jump unpredictably between peaks during acquisition.

The narrow peak of the fluorescence beads-derived sharpness function and its unimodal nature made the fluorescent beads well suited for benchmarking the algorithm, as even small deviations from the optimal focus caused noticeable changes in the sharpness metric. Furthermore, the peak position aligned with the sharpest image as judged visually by the authors.

Using the 20x objective and with the iPFS algorithm running in the background, we acquired 3D stacks of bead clusters at six different locations on the slide every 10 minutes over 40 h. The temperature in the incubator was maintained at $\qty{37}{\celsius}$ for the entire duration of the experiment to simulate conditions typical for live-sample imaging. 
The marker tracked by the algorithm was positioned slightly off-centre relative to the bead clusters (Figure S1). 

Figure \ref{fig:beads}D shows the Z positions of the sharpness maxima for the marker ($\zmark$, black line) and the beads ($z_{\rm beads}$, red line) at one of the six imaged locations over time. Both curves exhibit an upward trend; the initial and final positions differ by about \qty{30}{\um}. Without iPFS, the bead images would have gone out of focus within a few hours at this magnification. 
If iPFS fully compensated for drift, the difference $d=z_{\rm beads}-z_{\rm mark}$ should remain constant over time and equal to the (optical) thickness of the coverslip. To assess this, we plotted $\Delta z(t) = d(t)-d(0)$ in Figure \ref{fig:beads}E; perfect compensation would yield $\Delta z = 0$. Although $\Delta z$ shows slight temporal fluctuations, it stays within the width of the beads’ sharpness function $S_{\rm beads}$, indicating effective focus maintenance. The other five locations exhibited even greater stability (Figure \ref{fig:beads}F). 
Figure \ref{fig:beads}G shows the distribution of $\Delta z$ from all locations and time points. The FWHM of this distribution (\qty{1.5}{\um}) is comparable to theoretical  DOF of the 20x objective used here (\qty{2.2}{\um}). 

We can see from all these examples that the fluorescent beads sample is a good testbed for assessing the performance of our autofocus system.

\subsection{Comparison of iPFS with the hardware-based PFS}\label{ssec:ipfs_vs_pfs}

To enable a fair comparison between \emph{hardware} and \emph{software} focusing systems, we performed the same set of measurements using the same objectives as previously (20x and 40x). At each position and time point, we recorded Z-stacks over a range of $\pm z_\mathrm{range}$ around the best focal plane so that the resulting kymographs  from PFS and iPFS are directly comparable. Unlike iPFS, the test focal plane was determined by the hardware-based focusing system rather than by applying an offset $d(t)$, meaning that each position now has its own $d(t)$. This does not change the interpretation of the curves: if the PFS system functions perfectly, the beads’ sharpness profile would reach its maximum at $\Delta z(t) = 0$.

\begin{figure*}[!h]
    \centering
    \includegraphics[width=0.7\linewidth]{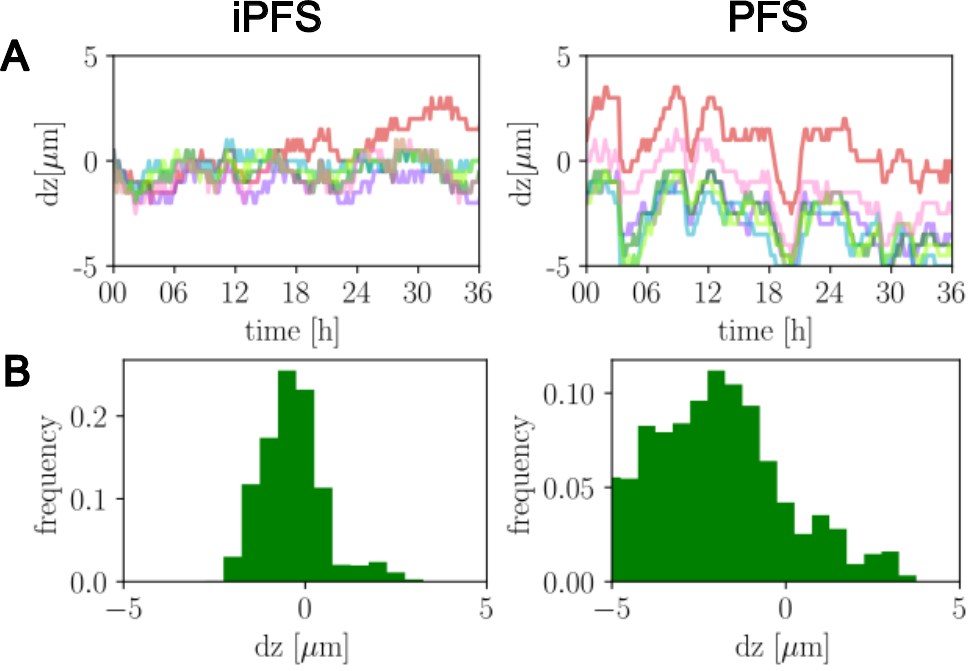}%
    \caption{Comparison between iPFS and PFS for the 20x objective. (A) Deviation between the position of the sharpest image of fluorescent beads and the position found by iPFS and PFS. (B) Histogram of $\Delta z$ (the same quantity as the one plotted in Figure \ref{fig:beads}G), for all positions and time points. }
    \label{fig:timeseries_ipfs_pfs_x20}
\end{figure*}

In Figure \ref{fig:timeseries_ipfs_pfs_x20}, we compare the two methods using the 20× objective by plotting $\Delta z$, defined as the difference between the focus position identified by each system and the ground-truth peak of the bead sharpness profile. iPFS displays consistently smaller deviations from this reference than PFS.

\begin{figure*}[!h]
    \centering
    \includegraphics[width=0.7\linewidth]{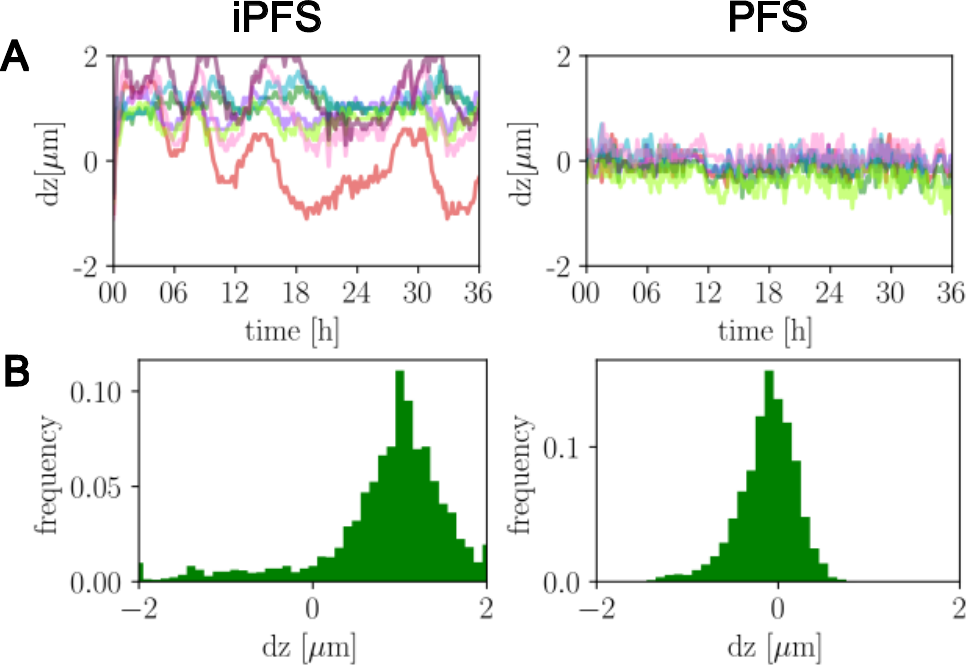}%
    \caption{Comparison between iPFS and PFS for the 40x objective. (A) The deviation between the position of the sharpest image of fluorescent beads and the position found by iPFS and PFS. (B) Histogram of $\Delta z$ for all positions and time points. }
    \label{fig:timeseries_ipfs_pfs_x40}
\end{figure*}

The situation is different with the 40x objective. Its much shallower depth of field makes it more sensitive to changes of the axial position, producing more peaked sharpness functions. As shown in Figure \ref{fig:timeseries_ipfs_pfs_x40}, both systems exhibit smaller overall fluctuations in $\Delta z$. However, the PFS performs substantially better with this objective: the sharpest planes at all positions vary only minimally, which is reflected in the narrow spread of $\Delta z$ values.
In contrast, iPFS shows only a modest improvement compared to its performance for the 20x objective. Although the spread of $\Delta z$ is slightly reduced, the time series display several large, correlated jumps. This suggests that the sample may have undergone physical changes that iPFS was not able to compensate for.

Table \ref{tab:focus_systems_bench} summarizes the results from Figures \ref{fig:timeseries_ipfs_pfs_x20} and \ref{fig:timeseries_ipfs_pfs_x40}.

\begin{table}[!h]
    \begin{tabular}{c c c c c }
      \toprule
      System                    & Objective   & $z_\mathrm{range} \,(\delta z)$      & FWHM of $\Delta z$ distribution                                & Focus quality \\
      \midrule
          \multirow{2}{*}{PFS}  
                                & 20x & not applicable    & \qty{4.1}{\um}   &poor                     \\
                                & 40x & not applicable  & \qty{0.4}{\um}  &very good                  \\
          \multirow{2}{*}{iPFS} 
                                & 20x & $\qty{5}{\micro\meter}\, (\qty{0.5}{\micro\meter})$   & \qty{1.5}{\um} &very good                 \\
                                & 40x & $\qty{2}{\micro\meter}\, (\qty{0.1}{\micro\meter})$   & \qty{0.6}{\um} &good                 \\
          \bottomrule
    \end{tabular}
    
    \caption{Comparison of iPFS and PFS for two objectives 20x and 40x. $z_{\rm range}$ and $\delta z$ refer to the scan range and scan step size of iPFS. FWHM of $\Delta z$ distribution represents the accuracy of tracking the sharpest focal plane. 
    }
    \label{tab:focus_systems_bench}
\end{table}

\subsection{Precautions}
\label{subsec:failure-modes}
\begin{figure}
  \centering
    \includegraphics[width=\columnwidth]{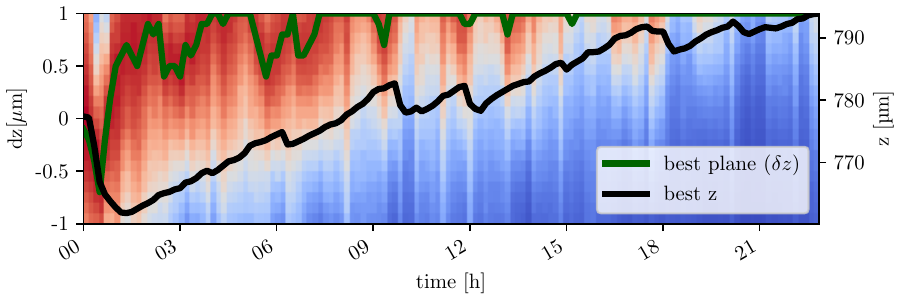}
    \caption{Example iPFS failure mode where the sharpest focus lies at the boundary of the scanned range. iPFS settings:  $\delta z = \qty{0.1}{\micro\meter}$ step, $\Delta z_\mathrm{focus} = \qty{1}{\micro\meter}$. The sharpest plane (green line for relative coordinates and solid black line for absolute coordinates) almost always lies at the edge of the Z-stack range which result in always defocused images.}
    \label{fig:ipfs_failure}
\end{figure}

As with any autofocus method, there are scenarios in which iPFS may not fully maintain the optimal focal plane. An illustrative example is shown in Figure \ref{fig:ipfs_failure}. In this case, although iPFS continues to adjust the focus upward over time (black line), the sharpest focal plane (yellow curve) eventually reaches the edge of the configured scanning range. This suggests that the true optimum moved too fast and eventually fell out of the scan range, and that the system was operating near the limits of the chosen acquisition parameters.

Importantly, this represents a deliberately challenging test condition: iPFS was configured with a very fine step size $(dz = \qty{0.1}{\micro\meter})$ and a shallow scanning range $(z_\mathrm{range} = \qty{1}{\micro\meter})$, prioritising precision over adaptability. In practical use, such situations can be easily mitigated by increasing the step size to enable faster tracking, or by expanding the scanning range to accommodate larger drifts without sacrificing performance. This example therefore highlights not a limitation of the method, but the importance of parameter tuning to match expected sample dynamics.

The performance of our method is mainly constrained by the speed with which the microscope hardware can complete the $Z$-scan and the time it takes to complete the actual image acquisition, during which iPFS is not active. If image acquisition takes little time (e.g., only one FOV, a single plane or a small Z stack), the effective limit on correctable drift is set by the scan range required to locate the global maximum of the sharpness function, together with the time needed to complete this scan. Since the $Z$-stage of modern microscopes moves at speeds of at least a few micrometres per second or faster (especially for piezo-driven stages), the bottlenecks are the settling time of the $Z$-stage and the communication speed between the computer, stage, and camera. For example, assuming the time of $300$ ms per frame (typical for our Nikon 2 Eclipse, including all these bottlenecks), and 20 frames per $Z$-stack of $10$ \mum height (step size $0.5$ \mum), axial drift cannot be larger than $(10\mum /2)/(20\times 300\mathrm{ms}) = 0.83\mum$/s $=3$ mm/h, which would be an exceedingly large drift. Such a large drift is unlikely to be of thermal origin, but it could arise in microfluidic experiments in which pressure changes inside the microfluidic channel cause significant distortions of the cover slip to which the device is bonded.

The performance gets worse if image acquisition during which iPFS is disabled takes a long time due to multiple positions, large Z-stacks, or multi-channel imaging. When acquisition time $T$ is the limiting factor ($T\gg$ the time to acquire an iPFS Z stack), the maximum correctable axial drift is $z_{\rm range}/T$, where $z_{\rm range}$ is the iPFS scan range as defined in Section \ref{sec:algorithm} (half of the total Z stack height). For example, if acquisition takes 1 minute and the iPFS scan range is $\pm 10$\mum, the maximum permitted drift velocity is $10$\mum/min $=0.6$ mm/h.

Tracking can also fail if the marker’s sharpness profile deviates from the ideal shown in Figure \ref{fig:16_dots}G. This is easily prevented with careful preparation:
\begin{itemize}
    \item Avoid image overexposure when imaging the marker, as image saturation can distort the sharpness metric. Our advice is to focus on the marker, find an empty region next to it, and adjust illumination and exposure time such that the region's pixel intensity lies approximately in the middle of the range of pixel values. 
    \item Ensure the marker remains stable over time. For example, when using oil-immersion objectives, the immersion oil may gradually degrade the ink and eat away finer details of the marker, affecting the sharpness profile to the point where the iPFS will stop working (SI Figure \ref{fig:SI_mark_w_bubbles}). To avoid this, the marker can be put on the sample side of the coverslip to prevent marker degradation, but this can contaminate the sample so it must be done with care.
    \item Nevertheless, as long as it is practical, do not put the marker on the sample side of the coverslip. If for some reason this cannot be prevented, care must be taken to ensure that the mark is far away from any motile or growing objects to avoid any interference with tracking (compare SI Figure \ref{fig:SI_precautions}).  
    \item Position the marker close — but not too close — to the region(s) of interest, to reduce the chance that moving objects enter the field of view and introduce competing sharpness peaks at different $z$ offsets.
    \item Secure the sample by firmly clamping it to prevent mechanical drift or loss of the marker from the field of view. 
\end{itemize}

Regarding the possibility of contaminating immersion objectives by debris from the marker, the marks created by common marker pens can be easily removed by isopropyl alcohol, which is advocated by many microscopists as a safe way of cleaning objectives \cite{duke_lens_2004}.

Lastly, the sharpness function we used does not account for varying illumination intensity, i.e., $S$ is not divided by average pixel intensity. We checked that normalizing by average intensity did not improve the performance of the algorithm, presumably because illumination does not significantly change during the relatively short acquisition of the sharpness metric. If unstable illumination is a problem, normalization should be added as follows:
\begin{equation}
  S = \sum_{x,y} \left( \sum_{i,j} R_{ij} I_{x+i,y+j}/I_{\rm mean} \right)^2.
\end{equation}
where $I_{\rm mean}$ is the mean pixel intensity of the image with the marker.

\subsection{Possible improvements to the iPFS}
While iPFS performs reliably in conditions investigated here, certain experimental scenarios may be challenging. Fortunately, with a few simple adjustments, the system can be made more robust: 
\begin{paragraph}{Surface triangulation:}
    Mechanical instabilities, such as improper clamping, can introduce motions beyond simple uniaxial translation, including sample rotation. To compensate, a single target dot can be replaced with at least three dots, enabling triangulation of the imaging plane (assuming it remains planar). This approach is implemented in the script \texttt{3\_point\_Z\_interpolation.bsh} provided in the software repository (Section \ref{ssec:data_availability}). 
    Thermally-induced bending can also make the reference surface non-planar. In such cases, three-point triangulation is insufficient, and a larger set of points distributed across the deformation region is required.
\end{paragraph}

\begin{paragraph}{Coupling with PFS:} The iPFS and PFS are not mutually exclusive. The PFS system can struggle when the sample undergoes large displacements—for example, after long intervals between image acquisitions (e.g., to reduce photobleaching) or if the sample is temporarily removed and then reinserted. In \mumgr, PFS-based autofocus always returns the stage to the previously recorded position before activating the PFS. If significant Z-drift has occurred in the meantime, the PFS may fail to find the interface and will switch off. In contrast, iPFS can maintain tracking during downtime by monitoring the air–glass interface where the markers are located, ensuring that the PFS starts from an already updated location.
\end{paragraph}

\paragraph{XY shift} A natural extension is to correct for XY shifts in addition to the Z-axis shift, which would account for three-dimensional translations. This approach is effective only if the marker and sample are part of the same rigid body, which may not be true for biological samples that are not attached to the coverslip.

\section{Application: imaging of \textit{Escherichia Coli} micro-colonies}\label{sec:application} 
To demonstrate the practical utility of our system, we applied it to time-lapse imaging of \ecoli{} bacteria. These cells are spherocylinders approximately \qty{1}{\micro\meter} in width, with an aspect ratio between 2 and 5 depending on their growth phase. Obtaining consistently sharp images is crucial for downstream analyses such as segmentation and cell tracking. 
We prepared an agarose slide as described in Section \ref{sec:fluor_iPFS} but instead of fluorescent beads, we inoculated a \qty{1}{\micro\liter} droplet of exponentially growing \ecoli{} culture with a low optical density $\approx 10^{-4}$, ensuring that primarily isolated, well-separated cells were present at the start of imaging. Images were acquired using the same 40x objective as before after placing the sample on the microscope's XY stage and turning on the incubator ($\qty{37}{\celsius}$). Without autofocus, the combined effect of bacterial growth and thermal drift caused by the incubator would quickly push the sample out of focus.

To see whether iPFS would manage to keep the sample in focus, the algorithm was configured with a step size $dz = \qty{0.1}{\um}$ and a scanning range of $\Delta z_\mathrm{focus} = \qty{3}{\um}$. 
To quantify the performance of iPFS,
bright-field Z-stacks were acquired with a \qty{1}{\micro\meter} step size, over a $\pm$\qty{5}{\micro\meter} range. The step size was slightly larger than the objective’s depth of field (\qty{0.7}{\um}). Such stacks can support high-accuracy segmentation using machine-learning approaches \cite{talissman}.
At each time point, a single GFP image was captured at the mid-plane ($\Delta z = 0$), both to verify accurate tracking of the focal plane and because fluorescence serves as a standard imaging modality.

\begin{figure*}
    \centering
    \includegraphics[width=1.0\linewidth]{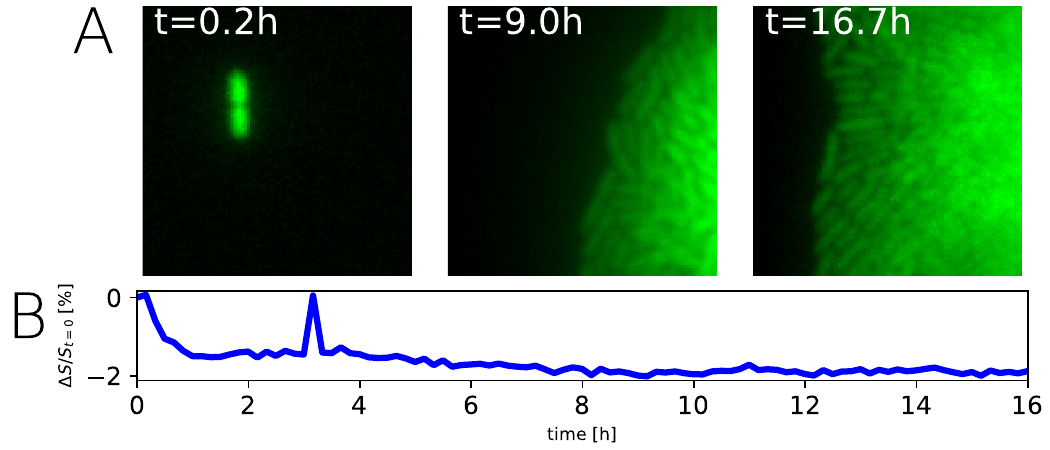}
    \caption{Practical application of iPFS: imaging a growing bacterial microcolony. (A) Images (left to right) show the colony-founding bacterium (about to divide) and the edge of the resulting colony at two later time points. Individual cells are clearly visible at the edge. Cells located farther from the edge form a multilayered structure and can no longer be distinguished. (B) The relative sharpness of the marker used by the iPFS as a function time. The sharpness changes very little, indicating good stability of iPFS.}
    
    \label{fig:application}
\end{figure*}

Figure \ref{fig:application}A shows that the image of the bacterial colony remained sharp over the 16 h of live imaging. The sharpness metric of the marker stayed within 2\% its initial value (Figure \ref{fig:application}B). We conclude that iPFS performed well for this challenging long-term imaging. This suggests that the iPFS may be suitable for other biological applications requiring stable, cost-effective autofocus.

Microcolonies of \emph{E. coli} are also a good test bed for higher magnification, immersion-based objectives. SI Figure \ref{fig:focus_maintained_100x} shows that iPFS can maintain focus on a growing bacterial colony using a 100x, 1.4 NA oil immersion objective.

\section{Comparison with other methods}
Compared to other approaches that use fiducial markers and software-based tracking \cite{colomb_estimation_2017,bian_autofocusing_2020}, our method combines several features that make it an attractive alternative:
\begin{itemize}
    \item our fiducial markers do not require special dyes, nanoparticles, or micropatterns; they can be created with a standard marker pen, making them essentially cost-free;
    \item the markers are placed on the side of the glass slide or plastic plate opposite the sample; they do not affect the biological sample, do not need to be sterile, and are not degraded through interaction with the sample;
    \item the algorithm does not rely on the biological sample itself for focusing, and is therefore applicable to samples lacking sharp features or undergoing significant changes over time;
    \item the algorithm used to locate (and focus on) the marker is simple, fast, and reliable; it does not require a powerful CPU or GPU and can be readily implemented with basic programming and microscopy expertise.
\end{itemize}

\section{Acknowledgements}
We thank Dr. Elin Lilja for kindly providing us with the \ecoli{} strain used in this research.

IDC acknowledges funding from the European Union's Horizon 2020 research and innovation programme under the Maria Skłodowska-Curie grant agreement No 847639. BW and IDC  acknowledge funding under Dioscuri, a programme initiated by the Max Planck Society, jointly managed with the National Science Centre in Poland, and mutually funded by the Polish Ministry of Science and Higher Education and German Federal Ministry of Education and Research (UMO-2019/02/H/NZ6/00003). 

This research was funded in part by National Science Centre, Poland, OPUS grant number 2021/43/B/NZ1/01436. For the purpose of Open Access, the authors have applied a CC-BY public copyright licence to any
Author Accepted Manuscript (AAM) version arising from this submission.

\section{Author contributions}
 BW conceptualized the system and wrote initial versions of the code. IDC further developed the code, designed benchmarks, performed experiments and analysed the data. Both authors contributed to writing the manuscript. 
 
\section{Materials and methods}
\subsection{Sample preparation}\label{sec:sample_preparation}
\paragraph{Colloidal beads.}
We used readily available, water-soluble acrylic fluorescent paint. After a 1:100 dilution in water, the suspension was centrifuged for 20 s at 3000 rpm. The supernatant, containing the smallest suspended particles, was collected by pipetting. Particle size was assessed microscopically, and most particles were significantly smaller than $\qty{2}{\micro\meter}$. 

The beads were deposited onto a 1 mm-thick agarose pad (2\% w/v) on a 1 mm glass slide and covered with a \#1.5 coverslip ($ \approx \qty{150}{\micro\meter}$ thickness). 
A schematic and representative image of this prepared benchmark slide are shown in  Figure \ref{fig:beads}A.

\paragraph{Markers.}
To enable reliable tracking of the glass surface, we use marks made with fine-tipped permanent black markers. We tried various models with success such as \texttt{Faber-Castell  Multimark permanent S} and \texttt{Q-CONNECT KF1134} with similar results (both mentioned models have a \qty{0.4}{\mm} tip size).
A very fine tip ensures that the mark left on the coverslip is small and does not obscure the entire field of view. The mark should be applied very gently, lifting the pen immediately upon contact with the glass. 

\subsection{Imaging setup}
\subsubsection{Microscopy\label{ssec:microscopy}}
Images were acquired using an automated Nikon Eclipse Ti2-E epi-fluorescent
microscope equipped with an automated XY stage and the Nikon Perfect Focus System version 4, and controlled by \mumgr  \cite{micromanager} version 2.0.1 (build dated 7th July 2022). To image fluorescent beads, we used an \texttt{ORCA-spark C11440-36U} (Hamamatsu, Japan). Fluorescent bacteria were imaged using a more sensitive \texttt{Andor Zyla 4.2 sCMOS camera}.

We used three different objectives:
\begin{itemize}
    \item $20\times$: \texttt{Nikon MRH10201 CFI Plan Fluor 20X}, NA 0.50, WD 2.1 mm, Ph1 DLL 
    \item $40\times$: \texttt{Nikon MRD70470 CFI Plan Apochromat Lambda 40X}, NA 0.95, WD 0.25-0.17mm , DIC N2
    \item $100\times$ \texttt{Nikon MRD01905 Plan Apo Lambda 100X}, NA 1.45, oil, WD 0.13mm, DIC N2
\end{itemize} 

Fluorescence imaging was performed using the filter cube \texttt{Semrock brightline FITC 2024B-NTE} (excitation: 470-500 \si{\nano\meter}, emission: 510-540 \si{\nano\meter}).
For illumination, we used CoolLED $\mathrm{pE-300^{white}\, SB}$ light source. Only the blue channel was enabled (emission peak $\approx \qty{450}{\nano\meter})$ and the power was set to 1\%, corresponding to $\approx$ \qty{0.7}{\mW} (40x objective) and \qty{0.8}{\mW} (20x objective) in the focal plane, as measured by Thorlabs PM16-130 optical power meter. This was sufficient to obtain strong fluorescence from both the beads and the bacteria. We did not measure the power for the 100x objective because it was not used for fluorescence imaging. 

\subsubsection{Computation of the FWHM}
In order to estimate the FWHM of the distributions, we replaced the discrete distribution of $\Delta z$ by continuous distributions obtained using the Kernel Density Estimate method \cite{parzen1962estimation}. We used $\sigma=0.5$ for the Gaussian kernel, which yielded good fits for all the distributions.

\subsubsection{Bacterial strains}

The \ecoli{} strain used in this article is EEL13 (MG1655 pA1\_gfp), genetically modified to constitutively express high levels of GFP. The strain (donated by Dr Elin Lilja) has been obtained from K-12 MG1655 by plasmid mediated gene replacement to insert constitutively expressed GFP in intergenic region between {\it pstS} and {\it glmS} genes. Cells were grown overnight at \qty{37}{\celsius} in M9 medium supplemented with 0.2\% glucose (v/v) and trace elements to $OD\geq 1$ and then diluted to OD$=10^{-4}$. The starting culture was therefore stationary.

\subsection{Data and code availability}\label{ssec:data_availability}
The \texttt{beanshell} script that can be run in \mumgr{}, along with the data that was used to generate the figures of this publication is available on \url{https://github.com/Dioscuri-Centre/iPFS}  

\subsection{Use of LLMs}
ChatGPT was used to improve clarity, conciseness, and stylistic consistency, while preserving the original scientific meaning. No LLMs were used to analyse the data, interpret the results, or to generate computer code.

\bibliographystyle{apsrev4-2}
\bibliography{bib_ipfs}

\providecommand{\noopsort}[1]{}\providecommand{\singleletter}[1]{#1}%
\begin{thebibliography}{29}%
\makeatletter
\providecommand \@ifxundefined [1]{%
 \@ifx{#1\undefined}
}%
\providecommand \@ifnum [1]{%
 \ifnum #1\expandafter \@firstoftwo
 \else \expandafter \@secondoftwo
 \fi
}%
\providecommand \@ifx [1]{%
 \ifx #1\expandafter \@firstoftwo
 \else \expandafter \@secondoftwo
 \fi
}%
\providecommand \natexlab [1]{#1}%
\providecommand \enquote  [1]{``#1''}%
\providecommand \bibnamefont  [1]{#1}%
\providecommand \bibfnamefont [1]{#1}%
\providecommand \citenamefont [1]{#1}%
\providecommand \href@noop [0]{\@secondoftwo}%
\providecommand \href [0]{\begingroup \@sanitize@url \@href}%
\providecommand \@href[1]{\@@startlink{#1}\@@href}%
\providecommand \@@href[1]{\endgroup#1\@@endlink}%
\providecommand \@sanitize@url [0]{\catcode `\\12\catcode `\$12\catcode
  `\&12\catcode `\#12\catcode `\^12\catcode `\_12\catcode `\%12\relax}%
\providecommand \@@startlink[1]{}%
\providecommand \@@endlink[0]{}%
\providecommand \url  [0]{\begingroup\@sanitize@url \@url }%
\providecommand \@url [1]{\endgroup\@href {#1}{\urlprefix }}%
\providecommand \urlprefix  [0]{URL }%
\providecommand \Eprint [0]{\href }%
\providecommand \doibase [0]{https://doi.org/}%
\providecommand \selectlanguage [0]{\@gobble}%
\providecommand \bibinfo  [0]{\@secondoftwo}%
\providecommand \bibfield  [0]{\@secondoftwo}%
\providecommand \translation [1]{[#1]}%
\providecommand \BibitemOpen [0]{}%
\providecommand \bibitemStop [0]{}%
\providecommand \bibitemNoStop [0]{.\EOS\space}%
\providecommand \EOS [0]{\spacefactor3000\relax}%
\providecommand \BibitemShut  [1]{\csname bibitem#1\endcsname}%
\let\auto@bib@innerbib\@empty
\bibitem [{\citenamefont {Kennard}\ \emph {et~al.}(2016)\citenamefont
  {Kennard}, \citenamefont {Osella}, \citenamefont {Javer}, \citenamefont
  {Grilli}, \citenamefont {Nghe}, \citenamefont {Tans}, \citenamefont
  {Cicuta},\ and\ \citenamefont
  {Cosentino~Lagomarsino}}]{kennard_individuality_2016}%
  \BibitemOpen
  \bibfield  {author} {\bibinfo {author} {\bibfnamefont {A.~S.}\ \bibnamefont
  {Kennard}}, \bibinfo {author} {\bibfnamefont {M.}~\bibnamefont {Osella}},
  \bibinfo {author} {\bibfnamefont {A.}~\bibnamefont {Javer}}, \bibinfo
  {author} {\bibfnamefont {J.}~\bibnamefont {Grilli}}, \bibinfo {author}
  {\bibfnamefont {P.}~\bibnamefont {Nghe}}, \bibinfo {author} {\bibfnamefont
  {S.~J.}\ \bibnamefont {Tans}}, \bibinfo {author} {\bibfnamefont
  {P.}~\bibnamefont {Cicuta}},\ and\ \bibinfo {author} {\bibfnamefont
  {M.}~\bibnamefont {Cosentino~Lagomarsino}},\ }\href
  {https://doi.org/10.1103/PhysRevE.93.012408} {\bibfield  {journal} {\bibinfo
  {journal} {Physical Review E}\ }\textbf {\bibinfo {volume} {93}},\ \bibinfo
  {pages} {012408} (\bibinfo {year} {2016})}\BibitemShut {NoStop}%
\bibitem [{\citenamefont {Duvernoy}\ \emph {et~al.}(2018)\citenamefont
  {Duvernoy}, \citenamefont {Mora}, \citenamefont {Ardré}, \citenamefont
  {Croquette}, \citenamefont {Bensimon}, \citenamefont {Quilliet},
  \citenamefont {Ghigo}, \citenamefont {Balland}, \citenamefont {Beloin},
  \citenamefont {Lecuyer},\ and\ \citenamefont
  {Desprat}}]{duvernoy_asymmetric_2018}%
  \BibitemOpen
  \bibfield  {author} {\bibinfo {author} {\bibfnamefont {M.-C.}\ \bibnamefont
  {Duvernoy}}, \bibinfo {author} {\bibfnamefont {T.}~\bibnamefont {Mora}},
  \bibinfo {author} {\bibfnamefont {M.}~\bibnamefont {Ardré}}, \bibinfo
  {author} {\bibfnamefont {V.}~\bibnamefont {Croquette}}, \bibinfo {author}
  {\bibfnamefont {D.}~\bibnamefont {Bensimon}}, \bibinfo {author}
  {\bibfnamefont {C.}~\bibnamefont {Quilliet}}, \bibinfo {author}
  {\bibfnamefont {J.-M.}\ \bibnamefont {Ghigo}}, \bibinfo {author}
  {\bibfnamefont {M.}~\bibnamefont {Balland}}, \bibinfo {author} {\bibfnamefont
  {C.}~\bibnamefont {Beloin}}, \bibinfo {author} {\bibfnamefont
  {S.}~\bibnamefont {Lecuyer}},\ and\ \bibinfo {author} {\bibfnamefont
  {N.}~\bibnamefont {Desprat}},\ }\href
  {https://doi.org/10.1038/s41467-018-03446-y} {\bibfield  {journal} {\bibinfo
  {journal} {Nature Communications}\ }\textbf {\bibinfo {volume} {9}},\
  \bibinfo {pages} {1120} (\bibinfo {year} {2018})}\BibitemShut {NoStop}%
\bibitem [{\citenamefont {Taheri-Araghi}\ \emph {et~al.}(2015)\citenamefont
  {Taheri-Araghi}, \citenamefont {Bradde}, \citenamefont {Sauls}, \citenamefont
  {Hill}, \citenamefont {Levin}, \citenamefont {Paulsson}, \citenamefont
  {Vergassola},\ and\ \citenamefont {Jun}}]{taheri-araghi_cell-size_2015}%
  \BibitemOpen
  \bibfield  {author} {\bibinfo {author} {\bibfnamefont {S.}~\bibnamefont
  {Taheri-Araghi}}, \bibinfo {author} {\bibfnamefont {S.}~\bibnamefont
  {Bradde}}, \bibinfo {author} {\bibfnamefont {J.~T.}\ \bibnamefont {Sauls}},
  \bibinfo {author} {\bibfnamefont {N.~S.}\ \bibnamefont {Hill}}, \bibinfo
  {author} {\bibfnamefont {P.~A.}\ \bibnamefont {Levin}}, \bibinfo {author}
  {\bibfnamefont {J.}~\bibnamefont {Paulsson}}, \bibinfo {author}
  {\bibfnamefont {M.}~\bibnamefont {Vergassola}},\ and\ \bibinfo {author}
  {\bibfnamefont {S.}~\bibnamefont {Jun}},\ }\href
  {https://doi.org/10.1016/j.cub.2014.12.009} {\bibfield  {journal} {\bibinfo
  {journal} {Current Biology}\ }\textbf {\bibinfo {volume} {25}},\ \bibinfo
  {pages} {385} (\bibinfo {year} {2015})}\BibitemShut {NoStop}%
\bibitem [{\citenamefont {Bakshi}\ \emph {et~al.}(2021)\citenamefont {Bakshi},
  \citenamefont {Leoncini}, \citenamefont {Baker}, \citenamefont
  {Cañas-Duarte}, \citenamefont {Okumus},\ and\ \citenamefont
  {Paulsson}}]{bakshi_tracking_2021}%
  \BibitemOpen
  \bibfield  {author} {\bibinfo {author} {\bibfnamefont {S.}~\bibnamefont
  {Bakshi}}, \bibinfo {author} {\bibfnamefont {E.}~\bibnamefont {Leoncini}},
  \bibinfo {author} {\bibfnamefont {C.}~\bibnamefont {Baker}}, \bibinfo
  {author} {\bibfnamefont {S.~J.}\ \bibnamefont {Cañas-Duarte}}, \bibinfo
  {author} {\bibfnamefont {B.}~\bibnamefont {Okumus}},\ and\ \bibinfo {author}
  {\bibfnamefont {J.}~\bibnamefont {Paulsson}},\ }\href
  {https://doi.org/10.1038/s41564-021-00900-4} {\bibfield  {journal} {\bibinfo
  {journal} {Nature Microbiology}\ }\textbf {\bibinfo {volume} {6}},\ \bibinfo
  {pages} {783} (\bibinfo {year} {2021})},\ \bibinfo {note} {number: 6
  Publisher: Nature Publishing Group}\BibitemShut {NoStop}%
\bibitem [{\citenamefont {Balagaddé}\ \emph {et~al.}(2005)\citenamefont
  {Balagaddé}, \citenamefont {You}, \citenamefont {Hansen}, \citenamefont
  {Arnold},\ and\ \citenamefont {Quake}}]{balagadde_long-term_2005}%
  \BibitemOpen
  \bibfield  {author} {\bibinfo {author} {\bibfnamefont {F.~K.}\ \bibnamefont
  {Balagaddé}}, \bibinfo {author} {\bibfnamefont {L.}~\bibnamefont {You}},
  \bibinfo {author} {\bibfnamefont {C.~L.}\ \bibnamefont {Hansen}}, \bibinfo
  {author} {\bibfnamefont {F.~H.}\ \bibnamefont {Arnold}},\ and\ \bibinfo
  {author} {\bibfnamefont {S.~R.}\ \bibnamefont {Quake}},\ }\href
  {https://doi.org/10.1126/science.1109173} {\bibfield  {journal} {\bibinfo
  {journal} {Science}\ }\textbf {\bibinfo {volume} {309}},\ \bibinfo {pages}
  {137} (\bibinfo {year} {2005})}\BibitemShut {NoStop}%
\bibitem [{\citenamefont {Wu}\ \emph {et~al.}(2013)\citenamefont {Wu},
  \citenamefont {Loutherback}, \citenamefont {Lambert}, \citenamefont
  {Estévez-Salmerón}, \citenamefont {Tlsty}, \citenamefont {Austin},\ and\
  \citenamefont {Sturm}}]{wu_cell_2013}%
  \BibitemOpen
  \bibfield  {author} {\bibinfo {author} {\bibfnamefont {A.}~\bibnamefont
  {Wu}}, \bibinfo {author} {\bibfnamefont {K.}~\bibnamefont {Loutherback}},
  \bibinfo {author} {\bibfnamefont {G.}~\bibnamefont {Lambert}}, \bibinfo
  {author} {\bibfnamefont {L.}~\bibnamefont {Estévez-Salmerón}}, \bibinfo
  {author} {\bibfnamefont {T.~D.}\ \bibnamefont {Tlsty}}, \bibinfo {author}
  {\bibfnamefont {R.~H.}\ \bibnamefont {Austin}},\ and\ \bibinfo {author}
  {\bibfnamefont {J.~C.}\ \bibnamefont {Sturm}},\ }\href
  {https://doi.org/10.1073/pnas.1314385110} {\bibfield  {journal} {\bibinfo
  {journal} {Proceedings of the National Academy of Sciences}\ }\textbf
  {\bibinfo {volume} {110}},\ \bibinfo {pages} {16103} (\bibinfo {year}
  {2013})}\BibitemShut {NoStop}%
\bibitem
  [{\url{https://www.microscopyu.com/applications/live-cell-imaging/correcting-focus-drift-in-live-cell-microscopy}()}]{drift-sources}%
  \BibitemOpen
  \url{https://www.microscopyu.com/applications/live-cell-imaging/correcting-focus-drift-in-live-cell-microscopy},\
  \href@noop {} {}\bibinfo {note} {Provides an interesting overview on the
  different sources of focus drift and correction methods}\BibitemShut
  {NoStop}%
\bibitem [{\citenamefont {Colomb}\ \emph {et~al.}()\citenamefont {Colomb},
  \citenamefont {Czerski}, \citenamefont {Sau},\ and\ \citenamefont
  {Sarkar}}]{colomb_estimation_2017}%
  \BibitemOpen
  \bibfield  {author} {\bibinfo {author} {\bibfnamefont {W.}~\bibnamefont
  {Colomb}}, \bibinfo {author} {\bibfnamefont {J.}~\bibnamefont {Czerski}},
  \bibinfo {author} {\bibfnamefont {J.}~\bibnamefont {Sau}},\ and\ \bibinfo
  {author} {\bibfnamefont {S.}~\bibnamefont {Sarkar}},\ }\href
  {https://doi.org/10.1111/jmi.12539} {\bibfield  {journal} {\bibinfo
  {journal} {Journal of Microscopy}\ }\textbf {\bibinfo {volume} {266}},\
  \bibinfo {pages} {298}},\ \bibinfo {note} {\_eprint:
  https://onlinelibrary.wiley.com/doi/pdf/10.1111/jmi.12539}\BibitemShut
  {NoStop}%
\bibitem [{\citenamefont {Firestone}\ \emph {et~al.}(1991)\citenamefont
  {Firestone}, \citenamefont {Cook}, \citenamefont {Culp}, \citenamefont
  {Talsania},\ and\ \citenamefont
  {Preston~Jr.}}]{firestoneComparisonAutofocusMethods1991}%
  \BibitemOpen
  \bibfield  {author} {\bibinfo {author} {\bibfnamefont {L.}~\bibnamefont
  {Firestone}}, \bibinfo {author} {\bibfnamefont {K.}~\bibnamefont {Cook}},
  \bibinfo {author} {\bibfnamefont {K.}~\bibnamefont {Culp}}, \bibinfo {author}
  {\bibfnamefont {N.}~\bibnamefont {Talsania}},\ and\ \bibinfo {author}
  {\bibfnamefont {K.}~\bibnamefont {Preston~Jr.}},\ }\href
  {https://doi.org/10.1002/cyto.990120302} {\bibfield  {journal} {\bibinfo
  {journal} {Cytometry}\ }\textbf {\bibinfo {volume} {12}},\ \bibinfo {pages}
  {195} (\bibinfo {year} {1991})}\BibitemShut {NoStop}%
\bibitem [{\citenamefont {Kristan}\ \emph {et~al.}(2006)\citenamefont
  {Kristan}, \citenamefont {Per{\v s}}, \citenamefont {Per{\v s}e},\ and\
  \citenamefont {Kova{\v c}i{\v
  c}}}]{kristanBayesspectralentropybasedMeasureCamera2006}%
  \BibitemOpen
  \bibfield  {author} {\bibinfo {author} {\bibfnamefont {M.}~\bibnamefont
  {Kristan}}, \bibinfo {author} {\bibfnamefont {J.}~\bibnamefont {Per{\v s}}},
  \bibinfo {author} {\bibfnamefont {M.}~\bibnamefont {Per{\v s}e}},\ and\
  \bibinfo {author} {\bibfnamefont {S.}~\bibnamefont {Kova{\v c}i{\v c}}},\
  }\href {https://doi.org/10.1016/j.patrec.2006.01.016} {\bibfield  {journal}
  {\bibinfo  {journal} {Pattern Recognition Letters}\ }\textbf {\bibinfo
  {volume} {27}},\ \bibinfo {pages} {1431} (\bibinfo {year}
  {2006})}\BibitemShut {NoStop}%
\bibitem [{\citenamefont {Ferzli}\ and\ \citenamefont
  {Karam}(2009)}]{ferzliNoReferenceObjectiveImage2009}%
  \BibitemOpen
  \bibfield  {author} {\bibinfo {author} {\bibfnamefont {R.}~\bibnamefont
  {Ferzli}}\ and\ \bibinfo {author} {\bibfnamefont {L.}~\bibnamefont {Karam}},\
  }\href {https://doi.org/10.1109/TIP.2008.2011760} {\bibfield  {journal}
  {\bibinfo  {journal} {IEEE Transactions on Image Processing}\ }\textbf
  {\bibinfo {volume} {18}},\ \bibinfo {pages} {717} (\bibinfo {year}
  {2009})}\BibitemShut {NoStop}%
\bibitem [{\citenamefont {Jia}\ \emph {et~al.}(2025)\citenamefont {Jia},
  \citenamefont {Ward}, \citenamefont {{van Tartwijk}}, \citenamefont {Yuan},
  \citenamefont {Feng},\ and\ \citenamefont
  {Kaminski}}]{jiaEnhancedMountainClimbing2025}%
  \BibitemOpen
  \bibfield  {author} {\bibinfo {author} {\bibfnamefont {Y.}~\bibnamefont
  {Jia}}, \bibinfo {author} {\bibfnamefont {E.~N.}\ \bibnamefont {Ward}},
  \bibinfo {author} {\bibfnamefont {F.~W.}\ \bibnamefont {{van Tartwijk}}},
  \bibinfo {author} {\bibfnamefont {Y.}~\bibnamefont {Yuan}}, \bibinfo {author}
  {\bibfnamefont {Y.}~\bibnamefont {Feng}},\ and\ \bibinfo {author}
  {\bibfnamefont {C.~F.}\ \bibnamefont {Kaminski}},\ }\href
  {https://doi.org/10.1088/2050-6120/ae008f} {\bibfield  {journal} {\bibinfo
  {journal} {Methods and Applications in Fluorescence}\ }\textbf {\bibinfo
  {volume} {13}},\ \bibinfo {pages} {045001} (\bibinfo {year}
  {2025})}\BibitemShut {NoStop}%
\bibitem [{\citenamefont {Shajkofci}\ and\ \citenamefont
  {Liebling}(2020)}]{shajkofciDeepFocusFewShotMicroscope2020}%
  \BibitemOpen
  \bibfield  {author} {\bibinfo {author} {\bibfnamefont {A.}~\bibnamefont
  {Shajkofci}}\ and\ \bibinfo {author} {\bibfnamefont {M.}~\bibnamefont
  {Liebling}},\ }in\ \href {https://doi.org/10.1109/ISBI45749.2020.9098331}
  {\emph {\bibinfo {booktitle} {2020 {{IEEE}} 17th {{International Symposium}}
  on {{Biomedical Imaging}} ({{ISBI}})}}}\ (\bibinfo {year} {2020})\ pp.\
  \bibinfo {pages} {164--168}\BibitemShut {NoStop}%
\bibitem [{\citenamefont {Luo}\ \emph {et~al.}(2021)\citenamefont {Luo},
  \citenamefont {Huang}, \citenamefont {Rivenson},\ and\ \citenamefont
  {Ozcan}}]{luoSingleShotAutofocusingMicroscopy2021}%
  \BibitemOpen
  \bibfield  {author} {\bibinfo {author} {\bibfnamefont {Y.}~\bibnamefont
  {Luo}}, \bibinfo {author} {\bibfnamefont {L.}~\bibnamefont {Huang}}, \bibinfo
  {author} {\bibfnamefont {Y.}~\bibnamefont {Rivenson}},\ and\ \bibinfo
  {author} {\bibfnamefont {A.}~\bibnamefont {Ozcan}},\ }\href
  {https://doi.org/10.1021/acsphotonics.0c01774} {\bibfield  {journal}
  {\bibinfo  {journal} {ACS Photonics}\ }\textbf {\bibinfo {volume} {8}},\
  \bibinfo {pages} {625} (\bibinfo {year} {2021})}\BibitemShut {NoStop}%
\bibitem [{\citenamefont {Grant}\ \emph {et~al.}(2014)\citenamefont {Grant},
  \citenamefont {Wac{\l}aw}, \citenamefont {Allen},\ and\ \citenamefont
  {Cicuta}}]{grantRoleMechanicalForces2014}%
  \BibitemOpen
  \bibfield  {author} {\bibinfo {author} {\bibfnamefont {M.~A.~A.}\
  \bibnamefont {Grant}}, \bibinfo {author} {\bibfnamefont {B.}~\bibnamefont
  {Wac{\l}aw}}, \bibinfo {author} {\bibfnamefont {R.~J.}\ \bibnamefont
  {Allen}},\ and\ \bibinfo {author} {\bibfnamefont {P.}~\bibnamefont
  {Cicuta}},\ }\bibfield  {journal} {\bibinfo  {journal} {Journal of The Royal
  Society Interface}\ }\href {https://doi.org/10.1098/rsif.2014.0400}
  {10.1098/rsif.2014.0400} (\bibinfo {year} {2014})\BibitemShut {NoStop}%
\bibitem [{ref({\natexlab{a}})}]{ref_PFS}%
  \BibitemOpen
  \href@noop {} {\bibinfo {title} {Nikon website}},\ \bibinfo {howpublished}
  {\url{https://www.microscopyu.com/applications/live-cell-imaging/nikon-perfect-focus-system}}
  ({\natexlab{a}})\BibitemShut {NoStop}%
\bibitem [{pgf({\natexlab{a}})}]{pgfocus_website}%
  \BibitemOpen
  \href@noop {} {\bibinfo {title} {pgfocus website}},\ \bibinfo {howpublished}
  {\url{https://web.archive.org/web/20240523145854/http://big.umassmed.edu/wiki/index.php/PgFocus}}
  ({\natexlab{a}})\BibitemShut {NoStop}%
\bibitem [{ref({\natexlab{b}})}]{ref_PFS2}%
  \BibitemOpen
  \href@noop {} {\bibinfo {title} {Nikon website}},\ \bibinfo {howpublished}
  {\url{https://www.microscopyu.com/tutorials/perfect-focus-offset-system-mechanics}}
  ({\natexlab{b}})\BibitemShut {NoStop}%
\bibitem [{pgf({\natexlab{b}})}]{pgfocus_hw}%
  \BibitemOpen
  \href@noop {} {\bibinfo {title} {pgfocus}},\ \bibinfo {howpublished}
  {\url{https://github.com/kbellve/pgFocus-hardware}}
  ({\natexlab{b}})\BibitemShut {NoStop}%
\bibitem [{pgf({\natexlab{c}})}]{pgfocus_table}%
  \BibitemOpen
  \href@noop {} {\bibinfo {title} {Table comparing pgfocus with other hardware
  systems}},\ \bibinfo {howpublished}
  {\url{https://web.archive.org/web/20240523145854/http://big.umassmed.edu/wiki/index.php/PgFocus\#Description}}
  ({\natexlab{c}})\BibitemShut {NoStop}%
\bibitem [{\citenamefont {Mir}\ \emph {et~al.}(2014)\citenamefont {Mir},
  \citenamefont {Xu},\ and\ \citenamefont {{van
  Beek}}}]{mirExtensiveEmpiricalEvaluation2014}%
  \BibitemOpen
  \bibfield  {author} {\bibinfo {author} {\bibfnamefont {H.}~\bibnamefont
  {Mir}}, \bibinfo {author} {\bibfnamefont {P.}~\bibnamefont {Xu}},\ and\
  \bibinfo {author} {\bibfnamefont {P.}~\bibnamefont {{van Beek}}},\ }in\ \href
  {https://doi.org/10.1117/12.2042350} {\emph {\bibinfo {booktitle}
  {{{IS}}\&{{T}}/{{SPIE Electronic Imaging}}}}},\ \bibinfo {editor} {edited by\
  \bibinfo {editor} {\bibfnamefont {N.}~\bibnamefont {Sampat}}, \bibinfo
  {editor} {\bibfnamefont {R.}~\bibnamefont {Tezaur}}, \bibinfo {editor}
  {\bibfnamefont {S.}~\bibnamefont {Battiato}},\ and\ \bibinfo {editor}
  {\bibfnamefont {B.~A.}\ \bibnamefont {Fowler}}}\ (\bibinfo {address} {{San
  Francisco, California, USA}},\ \bibinfo {year} {2014})\ p.\ \bibinfo {pages}
  {90230I}\BibitemShut {NoStop}%
\bibitem [{\citenamefont {Redondo}\ \emph {et~al.}(2012)\citenamefont
  {Redondo}, \citenamefont {Bueno}, \citenamefont {Valdiviezo}, \citenamefont
  {Nava}, \citenamefont {Crist{\'o}bal}, \citenamefont {D{\'e}niz},
  \citenamefont {{Garc{\'i}a-Rojo}}, \citenamefont {Salido}, \citenamefont
  {Fern{\'a}ndez}, \citenamefont {Vidal},\ and\ \citenamefont
  {{Escalante-Ram{\'i}rez}}}]{redondoAutofocusEvaluationBrightfield2012}%
  \BibitemOpen
  \bibfield  {author} {\bibinfo {author} {\bibfnamefont {R.}~\bibnamefont
  {Redondo}}, \bibinfo {author} {\bibfnamefont {G.}~\bibnamefont {Bueno}},
  \bibinfo {author} {\bibfnamefont {J.~C.}\ \bibnamefont {Valdiviezo}},
  \bibinfo {author} {\bibfnamefont {R.}~\bibnamefont {Nava}}, \bibinfo {author}
  {\bibfnamefont {G.}~\bibnamefont {Crist{\'o}bal}}, \bibinfo {author}
  {\bibfnamefont {O.}~\bibnamefont {D{\'e}niz}}, \bibinfo {author}
  {\bibfnamefont {M.}~\bibnamefont {{Garc{\'i}a-Rojo}}}, \bibinfo {author}
  {\bibfnamefont {J.}~\bibnamefont {Salido}}, \bibinfo {author} {\bibfnamefont
  {M.~D.~M.}\ \bibnamefont {Fern{\'a}ndez}}, \bibinfo {author} {\bibfnamefont
  {J.}~\bibnamefont {Vidal}},\ and\ \bibinfo {author} {\bibfnamefont
  {B.}~\bibnamefont {{Escalante-Ram{\'i}rez}}},\ }\href
  {https://doi.org/10.1117/1.JBO.17.3.036008} {\bibfield  {journal} {\bibinfo
  {journal} {Journal of Biomedical Optics}\ }\textbf {\bibinfo {volume} {17}},\
  \bibinfo {pages} {036008} (\bibinfo {year} {2012})}\BibitemShut {NoStop}%
\bibitem [{\citenamefont {Yang}\ \emph {et~al.}(2018)\citenamefont {Yang},
  \citenamefont {Berndl}, \citenamefont {Michael~Ando}, \citenamefont {Barch},
  \citenamefont {Narayanaswamy}, \citenamefont {Christiansen}, \citenamefont
  {Hoyer}, \citenamefont {Roat}, \citenamefont {Hung}, \citenamefont {Rueden},
  \citenamefont {Shankar}, \citenamefont {Finkbeiner},\ and\ \citenamefont
  {Nelson}}]{yangAssessingMicroscopeImage2018}%
  \BibitemOpen
  \bibfield  {author} {\bibinfo {author} {\bibfnamefont {S.~J.}\ \bibnamefont
  {Yang}}, \bibinfo {author} {\bibfnamefont {M.}~\bibnamefont {Berndl}},
  \bibinfo {author} {\bibfnamefont {D.}~\bibnamefont {Michael~Ando}}, \bibinfo
  {author} {\bibfnamefont {M.}~\bibnamefont {Barch}}, \bibinfo {author}
  {\bibfnamefont {A.}~\bibnamefont {Narayanaswamy}}, \bibinfo {author}
  {\bibfnamefont {E.}~\bibnamefont {Christiansen}}, \bibinfo {author}
  {\bibfnamefont {S.}~\bibnamefont {Hoyer}}, \bibinfo {author} {\bibfnamefont
  {C.}~\bibnamefont {Roat}}, \bibinfo {author} {\bibfnamefont {J.}~\bibnamefont
  {Hung}}, \bibinfo {author} {\bibfnamefont {C.~T.}\ \bibnamefont {Rueden}},
  \bibinfo {author} {\bibfnamefont {A.}~\bibnamefont {Shankar}}, \bibinfo
  {author} {\bibfnamefont {S.}~\bibnamefont {Finkbeiner}},\ and\ \bibinfo
  {author} {\bibfnamefont {P.}~\bibnamefont {Nelson}},\ }\href
  {https://doi.org/10.1186/s12859-018-2087-4} {\bibfield  {journal} {\bibinfo
  {journal} {BMC Bioinformatics}\ }\textbf {\bibinfo {volume} {19}},\ \bibinfo
  {pages} {77} (\bibinfo {year} {2018})}\BibitemShut {NoStop}%
\bibitem [{\citenamefont {Edelstein}\ \emph {et~al.}(2010)\citenamefont
  {Edelstein}, \citenamefont {Amodaj}, \citenamefont {Hoover}, \citenamefont
  {Vale},\ and\ \citenamefont {Stuurman}}]{micromanager}%
  \BibitemOpen
  \bibfield  {author} {\bibinfo {author} {\bibfnamefont {A.}~\bibnamefont
  {Edelstein}}, \bibinfo {author} {\bibfnamefont {N.}~\bibnamefont {Amodaj}},
  \bibinfo {author} {\bibfnamefont {K.}~\bibnamefont {Hoover}}, \bibinfo
  {author} {\bibfnamefont {R.}~\bibnamefont {Vale}},\ and\ \bibinfo {author}
  {\bibfnamefont {N.}~\bibnamefont {Stuurman}},\ }\href
  {https://doi.org/10.1002/0471142727.mb1420s92} {\bibfield  {journal}
  {\bibinfo  {journal} {Current Protocols in Molecular Biology}\ }\textbf
  {\bibinfo {volume} {92}},\ \bibinfo {pages} {14.20.1} (\bibinfo {year}
  {2010})}\BibitemShut {NoStop}%
\bibitem [{\citenamefont {Oldenbourg}\ and\ \citenamefont
  {Shribak}(2010)}]{Oldenbourg_Shribak}%
  \BibitemOpen
  \bibfield  {author} {\bibinfo {author} {\bibfnamefont {R.}~\bibnamefont
  {Oldenbourg}}\ and\ \bibinfo {author} {\bibfnamefont {M.}~\bibnamefont
  {Shribak}},\ }\bibinfo {title} {Microscopes}\ (\bibinfo {year} {2010})\ pp.\
  \bibinfo {pages} {28.1--28.62}\BibitemShut {NoStop}%
\bibitem [{\citenamefont {Duke}(2004)}]{duke_lens_2004}%
  \BibitemOpen
  \bibfield  {author} {\bibinfo {author} {\bibfnamefont {C.}~\bibnamefont
  {Duke}},\ }\href {https://doi.org/10.1017/S1551929500052007} {\bibfield
  {journal} {\bibinfo  {journal} {Microscopy Today}\ }\textbf {\bibinfo
  {volume} {12}},\ \bibinfo {pages} {42} (\bibinfo {year} {2004})}\BibitemShut
  {NoStop}%
\bibitem [{tal()}]{talissman}%
  \BibitemOpen
  \href@noop {} {\bibinfo {title} {Talissman}},\ \bibinfo {howpublished}
  {\url{https://github.com/jeanollion/TaLiSSman/}}\BibitemShut {NoStop}%
\bibitem [{\citenamefont {Bian}\ \emph {et~al.}()\citenamefont {Bian},
  \citenamefont {Guo}, \citenamefont {Jiang}, \citenamefont {Zhu},
  \citenamefont {Wang}, \citenamefont {Song}, \citenamefont {Zhang},
  \citenamefont {Hoshino},\ and\ \citenamefont
  {Zheng}}]{bian_autofocusing_2020}%
  \BibitemOpen
  \bibfield  {author} {\bibinfo {author} {\bibfnamefont {Z.}~\bibnamefont
  {Bian}}, \bibinfo {author} {\bibfnamefont {C.}~\bibnamefont {Guo}}, \bibinfo
  {author} {\bibfnamefont {S.}~\bibnamefont {Jiang}}, \bibinfo {author}
  {\bibfnamefont {J.}~\bibnamefont {Zhu}}, \bibinfo {author} {\bibfnamefont
  {R.}~\bibnamefont {Wang}}, \bibinfo {author} {\bibfnamefont {P.}~\bibnamefont
  {Song}}, \bibinfo {author} {\bibfnamefont {Z.}~\bibnamefont {Zhang}},
  \bibinfo {author} {\bibfnamefont {K.}~\bibnamefont {Hoshino}},\ and\ \bibinfo
  {author} {\bibfnamefont {G.}~\bibnamefont {Zheng}},\ }\href
  {https://doi.org/10.1002/jbio.202000227} {\bibfield  {journal} {\bibinfo
  {journal} {Journal of Biophotonics}\ }\textbf {\bibinfo {volume} {13}},\
  \bibinfo {pages} {e202000227}}\BibitemShut {NoStop}%
\bibitem [{\citenamefont {Parzen}(1962)}]{parzen1962estimation}%
  \BibitemOpen
  \bibfield  {author} {\bibinfo {author} {\bibfnamefont {E.}~\bibnamefont
  {Parzen}},\ }\href@noop {} {\bibfield  {journal} {\bibinfo  {journal} {The
  annals of mathematical statistics}\ }\textbf {\bibinfo {volume} {33}},\
  \bibinfo {pages} {1065} (\bibinfo {year} {1962})}\BibitemShut {NoStop}%
\end{thebibliography}%

\newpage
\section*{SUPPLEMENTARY INFORMATION}

\setcounter{equation}{0}
\setcounter{figure}{0}
\setcounter{table}{0}
\makeatletter
\renewcommand{\theequation}{S\arabic{equation}}
\renewcommand{\thefigure}{S\arabic{figure}}
\renewcommand{\bibnumfmt}[1]{[S#1]}
\renewcommand{\citenumfont}[1]{S#1}

\section*{Comparison of different image filters}

\begin{figure}[!h]
    \centering
  \tikz\node[label={north west:{\large A}
    }] {\includegraphics[height=8cm]{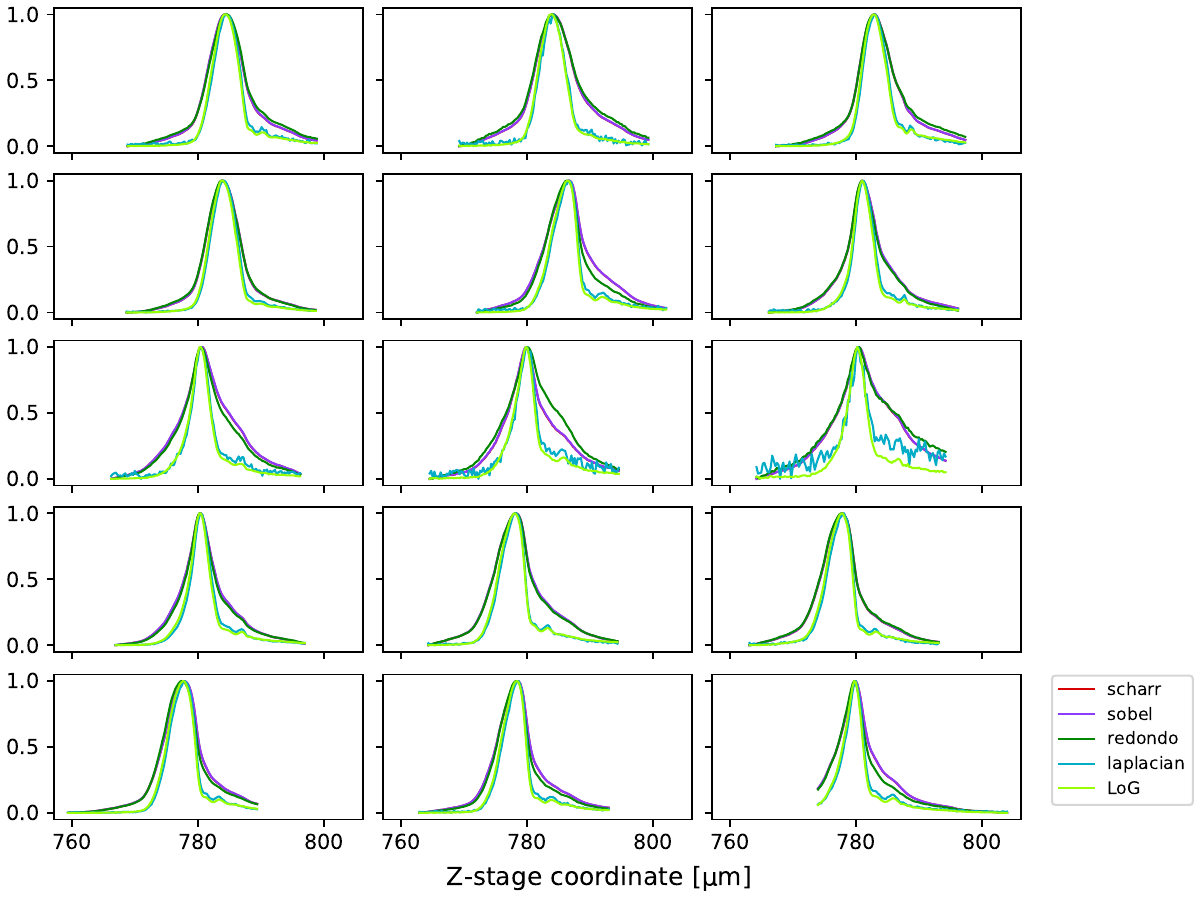}};%
  \tikz\node[label={north west:{\large B}}] {\includegraphics[height=8
  cm]{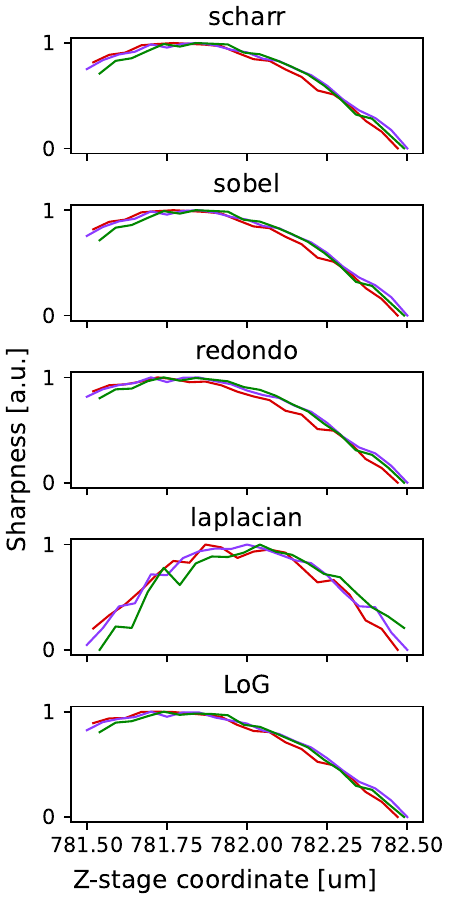}};
    \caption{(A) Comparison between different sharpness functions, obtained from 15 different markers (same as in Fig. \ref{fig:16_dots}), and five different filters. (B) Sharpness functions for a single dot imaged 3 times, for five different filters.
    }
    \label{fig:si_comparison_filters}
  \end{figure}

SI Figure \ref{fig:si_comparison_filters} compares the sharpness function calculated using different image filters: 

\begin{description}
    \item[\textbf{Redondo}]
    \[
    R =
        \begin{bmatrix}
             0 & 1 & 0  \\
             -3 & 0 & 1  \\
             0 & 1 & 0
        \end{bmatrix}.
    \]
    The sharpness function is calculated as in Eq. (2).

    \item[\textbf{Sobel} (horizontal/vertical gradient)]
    \[
        X = \begin{bmatrix}
          -1 & 0 & 1 \\
          -2 & 0 & 2 \\
          -1 & 0 & 1
        \end{bmatrix},
    \]
    and
    \[
    Y = \begin{bmatrix}
    -1 & -2 & -1 \\
     0 &  0 &  0 \\
     1 &  2 &  1
    \end{bmatrix}.
    \]
    The sharpness function is defined as 
    $S=\sum_{x,y} \left( \lvert \sum_{i,j}X_{ij}I_{x+i,y+j}\rvert+\lvert\sum_{i,j}Y_{ij}I_{x+i,y+j}\rvert \right)^2$ .
    \item[\textbf{Scharr} (horizontal/vertical gradient)]
    \[
    X = \begin{bmatrix}
    -3 & 0 & 3 \\
    -10 & 0 & 10 \\
    -3 & 0 & 3
    \end{bmatrix},
    \]
    and
    \[
    Y = \begin{bmatrix}
    -3 & -10 & -3 \\
     0 &  0 &  0 \\
     3 &  10 &  3
    \end{bmatrix}.
    \]
    The sharpness function is defined as 
    $S=\sum_{x,y} \left( \lvert \sum_{i,j}X_{ij}I_{x+i,y+j}\rvert+\lvert\sum_{i,j}Y_{ij}I_{x+i,y+j}\rvert \right)^2$ .

    \item[\textbf{Laplacian}]
    corresponds to applying twice the Sobel kernel in each direction, giving the following kernel:
    \[
     R=\begin{bmatrix}
    2 & 0 & 2 \\
    0 & -8 & 0 \\
    2 & 0 & 2 \\
    \end{bmatrix}.
    \]
    The sharpness function is calculated as in Eq. (2).
    \item[\textbf{Laplacian of Gaussian (LoG)}]
    Before using the Laplacian, the following blurring filter is applied to the source image:
    \[
    \begin{bmatrix}
    0.0625 & 0.125 & 0.0625 \\
    0.125 & 0.25 & 0.125 \\
    0.0625 & 0.125 & 0.0625
    \end{bmatrix}.
    \]
    Then the Laplacian filter is applied to the resulting image to obtain the sharpness function as described above.
\end{description}


Looking at Figure \ref{fig:si_comparison_filters}A, we note that
\begin{itemize}
    \item Different markers yield slightly different sharpness functions.
    \item Redondo, Scharr and Sobel filters yield very similar sharpness functions (with Scharr and Sobel almost indistinguishable), whereas Laplacian often results in a narrower peak.
    \item The Laplacian filter yields a more rugged sharpness function, with many local maxima, sometimes very close to the global maximum.
    \item Compared to the pure Laplacian filter, the LoG filter exhibits lower ruggedness. 
\end{itemize}
This could suggest that the LoG filter would be the preferred filter due to a sharp peak with a single maximum. However, it turns out that all these filters perform similarly when it comes to determining the position of maximum sharpness. To show this, we imaged the same marker repeatedly 3 times and compared the resulting sharpness functions (SI Fig. \ref{fig:si_comparison_filters}B), and the positions of maximum sharpness. We found that the standard errors of determining the position were $0.0118 \mum$ (Scharr), $0.0118 \mum$ (Sobel), $0.0054 \mum$ (Rendondo), $0.0241 \mum$ (Laplacian) and $0.0054 \mum$ (LoG). These values were very similar, suggesting that there was no clear advantage of one of the filters over another, and that Redondo (the filter we used in the final algorithm) performs at least as well as other filters.

\newpage
\section*{Supplementary figures}

\begin{figure}[!h]
    \centering
    \includegraphics[width=\columnwidth]{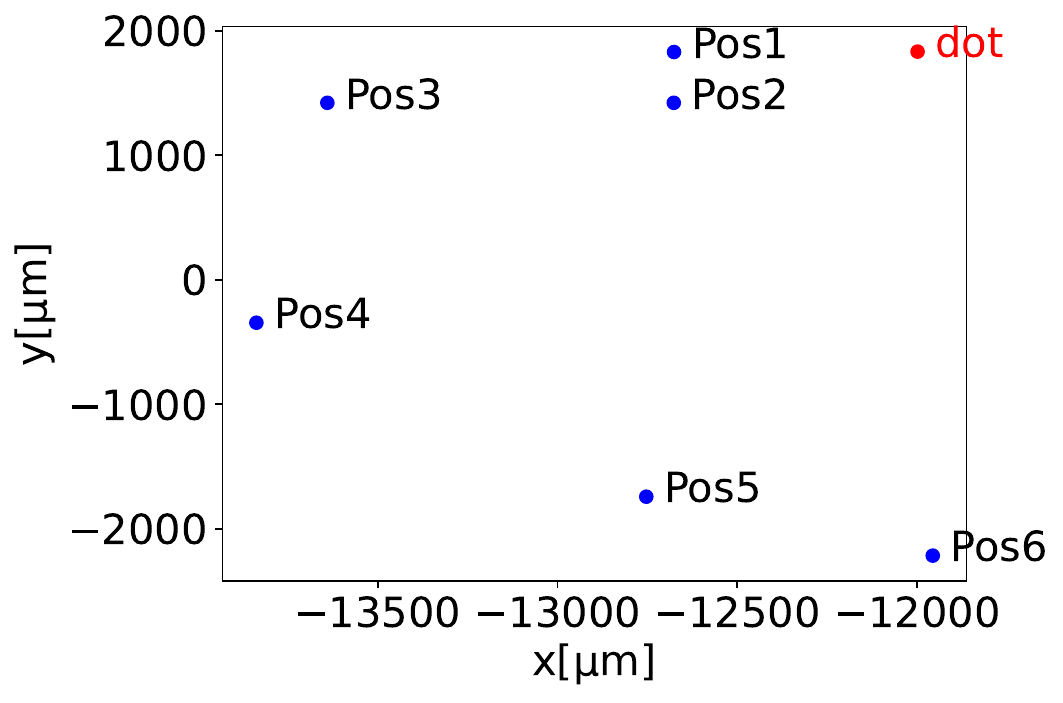}
    \caption{Positions of the imaged bead clusters (blue dots). The red dot marks the location used by iPFS to focus on the sample. The marker is placed on the opposite surface of the glass slide relative to the beads.}
    \label{fig:pos_x20}
\end{figure}

\begin{figure}[!h]
    \centering
    \includegraphics[width=0.5\linewidth]{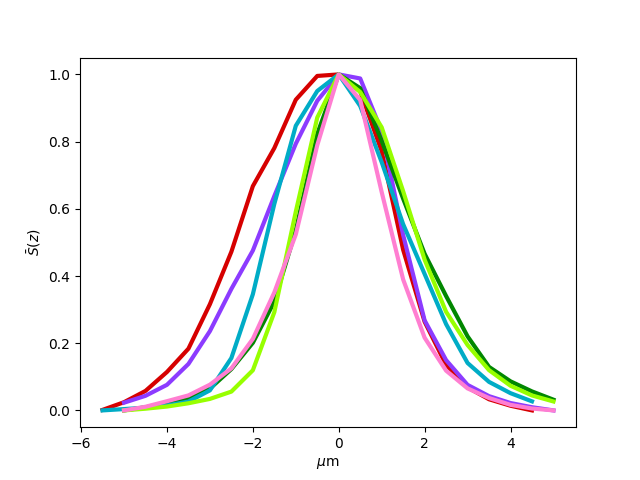}
    \caption{The sharpness function of the beads imaged in Fig. \ref{fig:timeseries_ipfs_pfs_x20} using the 20x objective.}
    \label{fig:sh_prof_dots_x20}
\end{figure}

\begin{figure}[!h]
    \centering
    \includegraphics[width=\linewidth]{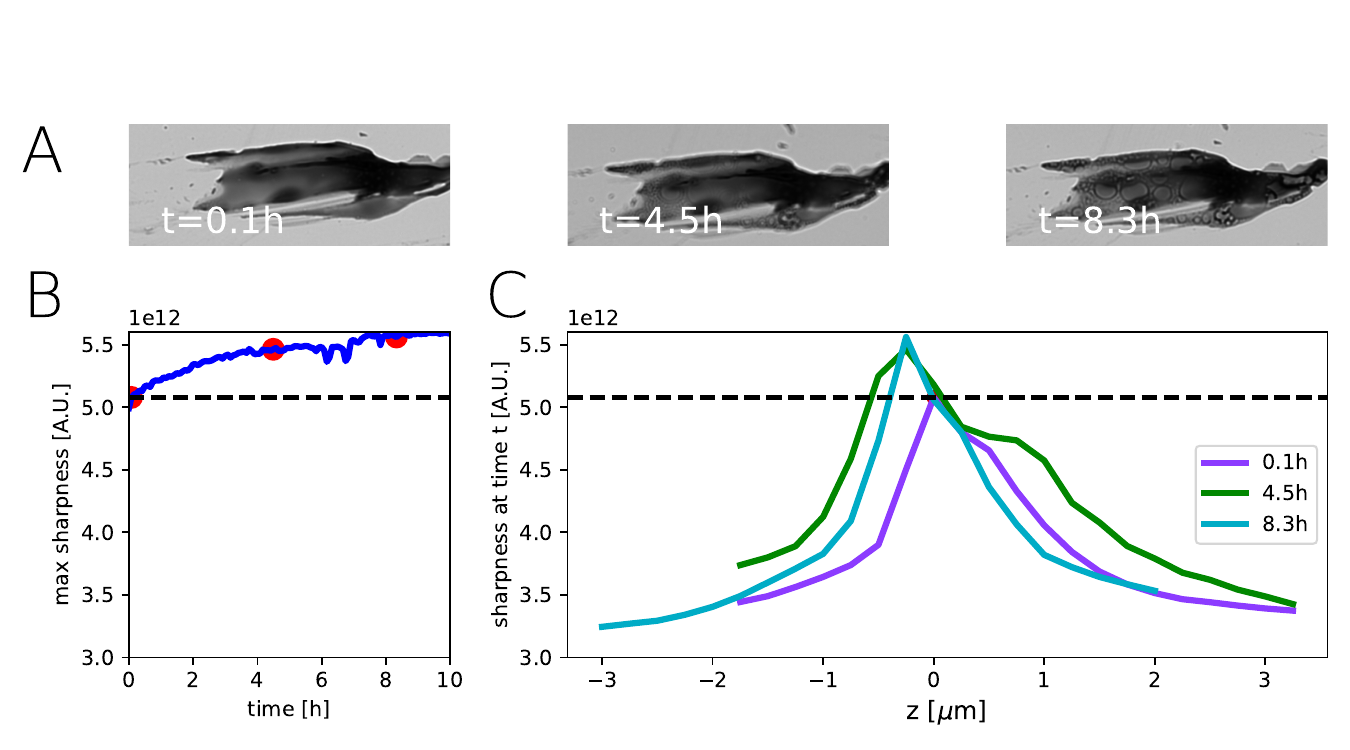}
    \caption{Marker degradation caused by immersion oil and its effect on the iPFS.  (A) The marker is attacked by the oil, which causes the formation of small bubbles over time. (B) The maximum of the sharpness function becomes more noisy as degradation worsens. Three points corresponding to images in panel A are marked in red. The reference sharpness when the  edges of the marker were correctly focused by iPFS and the marker was not degraded is indicated by dashed lines (C) Sharpness function for the three time points. As a result of marker degradation, the sharpness profile gains a higher maximum (visible in panel C) caused by the bubbles which also possess sharp edges, apparently sharper than the edges of the marker itself. This cause the iPFS to focus on the wrong object and cause iPFS to fail to maintain focus of imaged objects. }
    \label{fig:SI_mark_w_bubbles}
\end{figure}

\begin{figure}[!h]
    \centering
    \includegraphics[width=\linewidth]{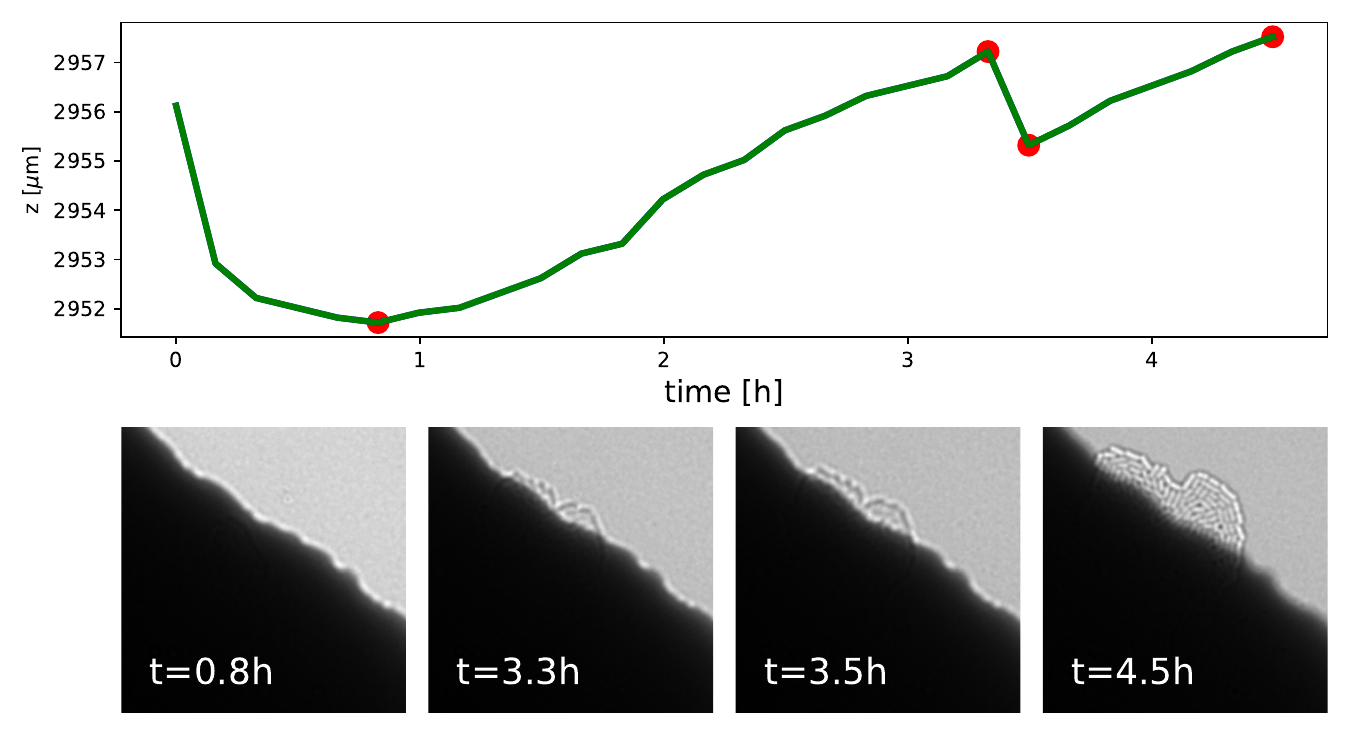}
    \caption{Example of iPFS failure due to bacterial growth interfering with the image of the marker. The marker was placed on the same side of the coverslip as a growing bacterial colony. As both the colony and the marker were in the same focal plane, the growing colony introduced a time-dependent contribution to the sharpness function, causing the algorithm to gradually shift its focus from the marker to the colony. Red dots in the plot (top) indicate the time each of the four images (bottom) was taken. }
    \label{fig:SI_precautions}
\end{figure}

\begin{figure}[!h]
    \centering
    \includegraphics[width=.5\linewidth]{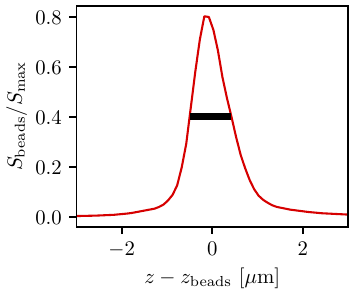}%
    \includegraphics[width=.5\linewidth]{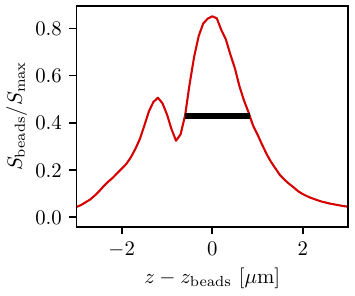}
    \caption{The averaged sharpness function of the beads from figure Fig. \ref{fig:beads}  across 4 locations in fluorescence (left) and bright-field (right), the FWHM is respectively $\approx$ \qty{0.9}{\um} and $\approx$ \qty{1.4}{\um}. Notice the secondary peak in the bright-field sharpness function. If this image ("sample") was used for autofocus, the autofocus algorithm could result in either of the two peaks being selected.}
    \label{fig:fwhm}
\end{figure}

\begin{figure}[!h]
    \centering
    \includegraphics[width=\linewidth]{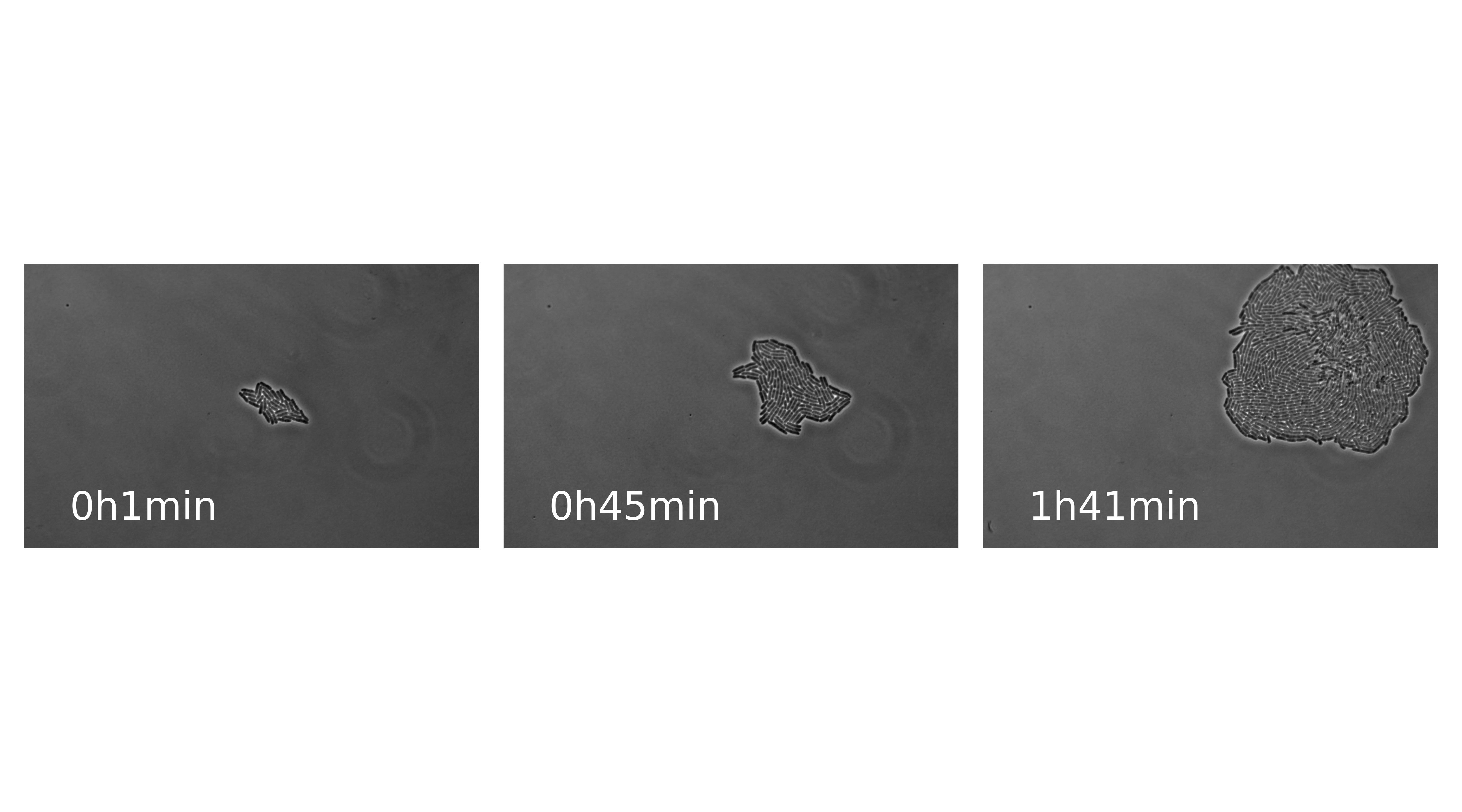}
    \caption{iPFS works also with high-NA oil objectives. A colony was imaged using a 100x oil objective (NA=1.4) for $\approx$ 2h. iPFS was able to maintain the focus for such duration. The right-hand side of the images is slightly out of focus due to a small sample tilt.
    }
    \label{fig:focus_maintained_100x}
  \end{figure}


\end{document}